# Testing a New Star Formation History Model from Principal Component Analysis to Facilitate Spectral Synthesis Modeling

Yanzhe Zhang[1], H. J. Mo[1], Katherine E. Whitaker[1], and Shuang Zhou[2]
[1] Department of Astronomy, University of Massachusetts, Amherst, MA 01003-9305, USA; yanzhezhang@umass.edu
[2] INAF—Osservatorio Astronomico di Brera, via Brera 28, I-20121 Milano, Italy

## Abstract

The spectrum of a galaxy is a complicated convolution of many properties of the galaxy, such as the star formation history (SFH), initial mass function, and metallicity. Inferring galaxy properties from the observed spectrum via spectral synthesis modeling is thus challenging. In particular, a simple yet flexible model for the SFH is required to obtain unbiased inferences. In this paper, we use SFHs from the IllustrisTNG and EAGLE simulations to test SFH models in terms of their capability of describing the simulated SFHs and the spectra generated from them. In addition to some commonly used SFH models ($\Gamma$, $\tau$, and nonparametric), we also examine a model developed from principal component analysis (PCA), trained by a set of SFHs from IllustrisTNG. We find that when using the first five principal components (eigenhistories), the PCA-based models can achieve a good balance between simplicity and accuracy. Among the models tested, the PCA-based model provides high flexibility, by capturing diverse and complex simulated SFHs. To accurately reproduce spectra generated from the simulated SFHs, it is necessary to have a degree of freedom to describe the most recent SFH (e.g., a step function covering the age of 0–0.3 Gyr). Overall, the PCA+step model performs well in capturing the diversity of SFHs and reproducing the associated spectra, suggesting it is a promising and reliable approach for spectral synthesis modeling.

## 1. Introduction

Observationally, many physical properties of a galaxy are encoded in its spectra and usually inferred through spectral synthesis modeling (SSM). However, such modeling is extremely challenging, because the main components, such as the star formation history (SFH), initial mass function (IMF), dust, and metallicity, are highly degenerate in the synthesized spectra (see the review by C. Conroy 2013). In particular, the SFH, one of the most crucial quantities of a galaxy, related to its assembly and evolution, is strongly degenerate in the spectral space, as each stellar population at a given age contributes to the spectrum over a large range of wavelength (e.g., D. Thomas et al. 2005; B. Lee et al. 2018).

With the advent of large and deep extragalactic surveys, the cosmic SFH, the average star formation rate (SFR) per comoving volume as a function of cosmic time (redshift, $z$), has become quite well established (P. Madau & M. Dickinson 2014). This history consists of a rising phase from high-$z$ to $z \sim 2$, with a gradually declining phase to the present time. However, the SFHs of individual galaxies are diverse. For example, the quenching of star formation may be driven by different mechanisms that depend on the stellar mass of galaxies (Y.-j. Peng et al. 2010), thereby giving rise to systematically different SFHs (e.g., A. Heavens et al. 2004; C. Pacifici et al. 2016). More recent studies of dwarf galaxies find that their SFHs are systematically different from those of galaxies of Milky Way mass (e.g., D. R. Weisz et al. 2011; G. Kauffmann 2014; Z. Lu et al. 2014, 2015; S. Zhou et al. 2020). The diversity of SFHs among the observed galaxy populations indicate that it is necessary to model the SFHs properly to make unbiased inferences from the observed spectra. Three qualities are needed for a successful model. First, the model should be simple, so that it can be constrained by observations. Second, the model should be flexible, so that it can cover the diversity in SFHs. Third, the model should be reliable, so that it does not lead to biased predictions from the spectra. The third requirement is essential, because the transformation from an SFH to a synthesized spectrum is highly nonlinear.

Various SFH models have been proposed for SSM (e.g., A. C. Carnall et al. 2019; J. Leja et al. 2019; K. A. Suess et al. 2022). Some commonly used parametric models include the $\Gamma$ model (Z. Lu et al. 2014), the exponentially declining ($\tau$ and delayed-$\tau$) model (B. Lee et al. 2018; P.-F. Wu et al. 2018; A. C. Carnall et al. 2019; S. Jain et al. 2024), the double-power-law model (P. S. Behroozi et al. 2013; M. D. Gladders et al. 2013; C. Pacifici et al. 2016; A. C. Carnall et al. 2018), and the log-normal model (M. D. Gladders et al. 2013). Nonparametric models based on binning the SFHs into histograms have also been used, either with fixed time bins (D. R. Weisz et al. 2011; J. Leja et al. 2019) or with variable time bins (K. A. Suess et al. 2022). Yet the applicability and performance of these models have not been tested systematically, according to the criteria described above, and it is still unclear which model is preferred in real applications (e.g., S. Zhou et al. 2020; D. Narayanan et al. 2024).

To better investigate the diversity of galaxy populations in their SFHs, we take advantage of hydrodynamic cosmological simulations that can predict the SFHs of individual galaxies. This will allow us to test directly the performance of different

SFH models, as well as the spectra predicted by these SFH models, in comparison to those obtained directly from the simulated SFHs. In this study, we use data from IllustrisTNG—a high-resolution cosmological hydrodynamic simulation of galaxy formation (D. Nelson et al. 2017; A. Pillepich et al. 2017a; V. Springel et al. 2017; F. Marinacci et al. 2018; J. P. Naiman et al. 2018). We select all central galaxies with stellar masses larger than $1 \times 10^9 h^{-1} M_\odot$ at $z = 0$ from the IllustrisTNG 100-1 simulation as our main sample. In addition to testing existing models, we also test a new set of models developed from principal component analysis (PCA); we refer to these models hereafter as PCA-based models. We apply PCA to the simulated SFHs to obtain a set of principal components (PCs; the eigenhistories) and use several lower-order eigenhistories as the base function to model the SFHs of galaxies. We demonstrate the flexibility and reliability of the PCA-based models by comparing them to existing models.

The paper is organized as follows. In Section 2, we introduce the training data set (IllustrisTNG) and the software (Bayesian Inference of Galaxy Spectra or BIGS) that generates the mock galaxy spectra. In Section 3, we introduce a few traditional SFH models we intend to test and describe our PCA-based models. In Section 4, we first test the SFH models in terms of their capacity to describe the input SFHs of individual galaxies in the simulated sample. We then generate synthesized spectra using the best-fit SFHs and compare them to spectra generated by the input SFHs. We also study how the predicted spectra are affected by the presence of very recent star formation and test how the acceptability of different models changes with the signal-to-noise ratio (SNR) of the spectra. In Section 5, we test the validity of the PCA-based models by applying them to galaxy samples of different properties and to galaxies from an independent hydro simulation. We further discuss and summarize our main conclusions in Section 6.

The cosmology model used in this paper is adopted from Planck Collaboration et al. (2016), which has $\Omega_{\Lambda,0} = 0.69$, $\Omega_{m,0} = 0.31$, $\Omega_{b,0} = 0.049$, and $H_0 = 100h \, \mathrm{km\,s^{-1}\,MPc^{-1}}$, with $h = 0.68$. This cosmology is also used in the TNG simulation.

## 2. Data and Software

### 2.1. Training Data

We use data from the publicly available IllustrisTNG project, which is a set of cosmological hydrodynamic simulations (D. Nelson et al. 2017; A. Pillepich et al. 2017a; V. Springel et al. 2017; F. Marinacci et al. 2018; J. P. Naiman et al. 2018). IllustrisTNG is built upon the moving mesh code AREPO (V. Springel 2010) and the success of the former Illustris galaxy formation model (S. Genel et al. 2014; M. Vogelsberger et al. 2014a, 2014b; D. Sijacki et al. 2015). Several improvements are implemented in IllustrisTNG, including a new model of feedback effects from supermassive black holes (R. Weinberger et al. 2016) and some improvements in the numerical framework (A. Pillepich et al. 2017b).

The IllustrisTNG project includes simulations in three cubic boxes, with comoving side lengths $L_{\rm box}$ roughly equal to 50 (TNG50), 100 (TNG100), and 300 (TNG300) Mpc, respectively. The largest simulation, TNG300, covers a relatively large physical volume and thus enables studies of galaxy clustering, rare and massive objects, and statistical analyses of galaxies. On the other hand, the smallest one, TNG50, offers a more detailed mass resolution of galaxies, allowing investigations of the structural properties of galaxies and small-scale gas processes in and around galaxies. TNG100 provides a balance between volume and mass resolution. We use the highest-resolution run of the TNG100 simulation, TNG100-1 (hereafter, TNG100), which simulates galaxy evolution from $z = 127$ to $z = 0$ in a box with $L_{\rm box} \sim 110.7$ Mpc ($75 \, \mathrm{cMpc\,h^{-1}}$). The minimum baryonic particle mass is $m_{\rm gas} \sim 1.4 \times M_\odot$.

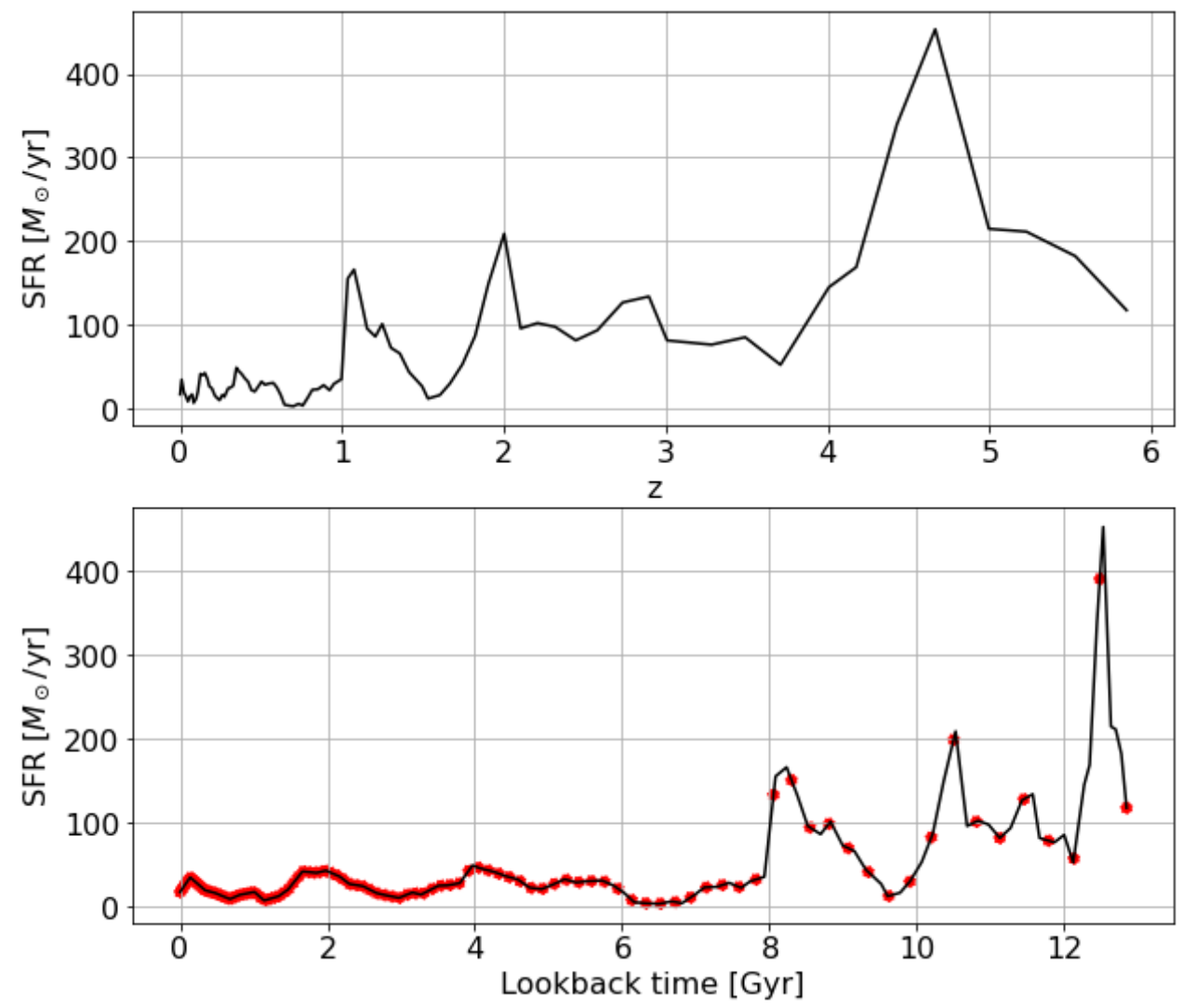

**Figure 1.** An example of the TNG100 SFH, sampled in the original redshift (upper panel) and in lookback time (lower panel). The solid black line is the original sampling in cosmic lookback time, and the red dots are the resampling in a logarithmic (10-based) time grid.

We select 16,141 galaxies at $z = 0$ with a stellar mass threshold $M_* = 1.0 \times 10^9 h^{-1} M_\odot$, thereby ensuring that these galaxies are well resolved in TNG100 at higher redshifts. The sample includes both central and satellite galaxies. We find that some of the satellite galaxies in the simulation have zero SFR at low-$z$, partly because of the quenching of star formation by halo-specific environmental effects and partly because of the numerical difficulties in following the evolution of satellite galaxies accurately. We therefore exclude satellite galaxies from the training data set, which gives a sample of 9168 central galaxies at $z = 0$. This choice not only reduces the potential impacts of numerical uncertainties in satellite galaxies, but also provides an opportunity to test whether or not our approach depends significantly on the training data adopted. As shown in Section 5.2, the PCA-based models trained by central galaxies are also valid for satellite galaxies. We use the instantaneous SFRs combined from the main and side branches of the subhalo merger tree rooted in the host subhalo of a galaxy. The SFRs do not take mass loss into account.

For each central galaxy at $z = 0$, a total of 100 snapshots of redshift from $z = 127$ to $z = 0$ are available. We use the SFHs only where $z \lesssim 6$, comprising 87 snapshots. We convert the redshift into cosmological time using the cosmological model adopted by the simulation. In developing the new PCA-based models in Section 3.2.2, we resample the SFH by interpolating the TNG100 data onto a 100-point time grid in logarithmic scale (base 10) that spans cosmic time up to $z \lesssim 6$. In Figure 1, we show one example to demonstrate our resampling of the SFH. The upper panel shows the original TNG100 sampling with respect to redshift, while the lower panel includes the converted cosmological lookback time (the black line) and our new interpolated sampling (the red dots) as a function of lookback time. Whereas the original sampling emphasizes older stellar populations at higher redshifts, our new sampling gives more weight to star formation at the most recent lookback times. More details can be found in Section 3.

### 2.2. Stellar Spectra

We generate the spectral energy distribution (SED) of each galaxy from its SFH using a stellar population synthesis code, BIGS, developed by S. Zhou et al. (2019). BIGS utilizes a full Bayesian analysis to fit the composite spectrum of a galaxy, along with several physical properties, such as the IMF, SFH, metallicity ($Z$), and dust attenuation. This code has been applied to MaNGA data to constrain the IMF in early-type galaxies (S. Zhou et al. 2019), to simultaneously model the mass accumulation and chemical evolution for spiral galaxies (S. Zhou et al. 2022), the SFHs of low-mass galaxies (S. Zhou et al. 2020), and massive red spiral galaxies (S. Zhou et al. 2021) in the local Universe. One of our eventual goals is to implement the new PCA-based models obtained in this study into BIGS. For the purpose of this paper, however, we only use BIGS to generate mock spectra from given SFHs. We refer the reader to the above references for a more detailed description of the code and the Bayesian inferences.

BIGS provides choices of different IMFs and templates of stellar spectra. To better isolate the effects produced by using different SFH models, we fix all parts other than the SFH, including adopting a Chabrier stellar IMF (G. Chabrier 2003) and the Padova 1994 isochrones (G. Bertelli et al. 1994), and test several SFH models, as detailed in Section 3. Combining the simulated SFHs with the BC03 simple stellar population templates (G. Bruzual & S. Charlot 2003), constructed with the STELIB stellar library, which has a wavelength range covering 3200–9500 Å with a spectral resolution of $\sim$3 Å (J. F. Le Borgne et al. 2003), and the Calzetti dust attenuation curve (D. Calzetti et al. 2000), we generate composite mock spectra for the simulated galaxies in our sample. We do not include broadening produced by stellar kinematics, but we test the effects of spectral resolutions and the SNR. Unless stated otherwise, we adopt a solar metallicity and an $E(B-V)$ value of 0.1.

## 3. SFH Models

In general, the SFH of a galaxy can be complicated, in the sense that it is hard to come up with a universal model for different galaxies. From the observational point of view, inferring the SFH of a galaxy from its spectrum is challenging, because of the degeneracy of the SFH with the IMF, metallicity, dust attenuation, and other factors. Consequently, inferences of SFHs from the observed spectra have to be made by adopting simple models for the functional form of the SFH. In this section, we first introduce three traditional SFH models: the $\Gamma$ model, the $\tau$ model, and the nonparametric stepwise model. We then present the development of our new SFH model based on PCA.

### 3.1. Traditional Models

Traditionally, the SFHs of galaxies are modeled by two approaches: (1) the parametric approach, which assumes a functional form characterized by a small number of parameters; and (2) the nonparametric approach, which models the SFH using histograms with various time bins. In this paper, we will test two parametric models, the $\Gamma$ model and $\tau$ model, and a stepwise model with a given number of time bins.

*The $\Gamma$ model*. This model is defined by a $\Gamma$ function, representing the SFR as a function of cosmic time $t$:

$$\Psi(t) = \frac{1}{\tau\gamma(\alpha,\, t_0/\tau)}\left(\frac{t_{\rm lb}}{\tau}\right)^{\alpha-1} e^{-t_{\rm lb}/\tau}, \tag{1}$$

where

$$t_{\rm lb} \equiv t_0 - t \tag{2}$$

is the lookback time, with $t_0$ being the present time, and

$$\gamma(\alpha,\, t_0/\tau) \equiv \int_0^{t_0/\tau} x^{\alpha-1} e^{-x} dx\ . \tag{3}$$

The model is thus specified by two free parameters, $\alpha$ and $\tau$, and normalized so that $\int_0^{t_0} \Psi(t)dt = 1$. The motivation for including the $\Gamma$ model comes from the empirical model of Z. Lu et al. (2014), who found that such a model can describe the SFHs of many galaxies in their empirical model of galaxy formation. Moreover, it was also implemented in BIGS to study different types of galaxies in the local Universe (S. Zhou et al. 2020, 2021). Given that there are only two free parameters, it is one of the simplest SFH models. Yet, for the same reason, this model may introduce biases in the inferences of physical parameters, because it is not very flexible (A. C. Carnall et al. 2019). We include this model in our study to test its applicability and for comparison with other models.

*The $\tau$ model*. This is an exponentially declining SFH, parameterized by the e-folding time $\tau$, and has the form

$$\Psi(t) = \begin{cases} \frac{1}{\tau\gamma(1,\,[t_0-t_{\rm s}]\,/\,\tau)} \exp\left(-\frac{t-t_{\rm s}}{\tau}\right), & t > t_{\rm s}; \\ 0, & \text{otherwise}, \end{cases} \tag{4}$$

where $t_{\rm s}$ is the specific time when star formation starts, from zero to the maximum value. Similar to the $\Gamma$ model, this model also has two free parameters, $t_{\rm s}$ and $\tau$. The $\tau$ model is one of the most commonly applied SFH models, because of its simplicity. Although recent improvements in statistical and computational techniques allow for more complex SFH models in SED fitting, the $\tau$ model is still widely used for comparison (e.g., B. Lee et al. 2018; P.-F. Wu et al. 2018; A. C. Carnall et al. 2019; S. Jain et al. 2024). However, this model seems to be disfavored by applications to high-redshift galaxies (N. A. Reddy et al. 2012) and in fitting simulated galaxies (A. C. Carnall et al. 2018; G. D. Joshi et al. 2021), as it can lead to significantly biased inferences (C. Pacifici et al. 2015; Y. Kaushal et al. 2024).

There are other parametric SFH models in the literature, such as the delayed-$\tau$, log-normal, and double-power-law models. A. C. Carnall et al. (2019) show that distinguishing between these parametric models using broadband photometry is hard. Since the global shapes of SFHs represented by these models are roughly covered by the two parametric models described above, we only use the $\Gamma$ and $\tau$ models in our study to represent this set of parametric models.

*The stepwise model*. Nonparametric models are found to be more flexible than parametric models in describing the diversity of SFH shapes. The accuracy of such a model depends on how many time bins are used in the model, which is a trade-off with computational tractability. Using an infinite number of time bins would return the "real" SFH, but this is unrealistic in relation to computational power and time. It is also unnecessary, as real observational data can only constrain a finite number of model parameters. J. Leja et al. (2019) demonstrate that at least five time bins are needed to carry the information from their mock photometric data. Yet, whether nonparametric models may possibly have hidden biases in SFH reconstructions via SSM is still an open question (S. Zibetti et al. 2024). Here, we adopt the same model as used in BIGS (S. Zhou et al. 2020, 2021), which is a slightly modified version of the model developed by D. R. Weisz et al. (2011). This model samples the average SFRs in seven fixed time intervals, specifically:

$$\begin{aligned} t &< 200\text{Myr} \\ 200 < t &< 500\text{Myr} \\ 500\text{Myr} < t &< 1\text{Gyr} \\ 1 < t &< 2\text{Gyr} \\ 2 < t &< 6\text{Gyr} \\ 6 < t &< 10\text{Gyr} \\ 10\text{Gyr} &< t. \end{aligned} \tag{5}$$

For this model, we follow S. Zhou et al. (2020, 2021) in normalizing the SFH, by setting SFR = 1 for the (1–2) Gyr bin and treating the average SFRs in the remaining six bins as free parameters. Once the relative heights of the SFRs are determined, the absolute SFRs are scaled up and down together, based on the total stellar mass. As a result, the stepwise model contains 6 degrees of freedom.

### 3.2. Models Based on PCA

#### 3.2.1. PCA

PCA is a commonly used method in data modeling. The general idea behind PCA is to identify patterns and relationships in high-dimensional data sets and then transform the original correlated high-dimensional coordinates into a set of new uncorrelated coordinates, in directions along which the data have the largest variances. These new coordinates, namely the PCs, are ranked in order of decreasing variance that the individual PCs account for. Therefore, a PCA of a data set allows us to identify a number of low-order PCs, to describe the most significant trends (features) of the entire data set, with higher-order PCs containing information about the details that are either uninteresting to us or cannot be constrained by data.

In astronomy, PCA has been employed in galaxy spectral classification (e.g., A. J. Connolly et al. 1995; C. W. Yip et al. 2004), identification between quiescent and star-forming galaxies (e.g., V. Wild et al. 2008), studies of galaxy clustering (e.g., M. Tegmark & B. C. Bromley 1999; S. Bonoli & U. L. Pen 2009; N. Hamaus et al. 2010; S. Zhou et al. 2023), the interstellar medium (e.g., H. Ungerechts et al. 1997; D. A. Neufeld et al. 2007; N. Lo et al. 2009; G. J. Melnick et al. 2011; P. A. Jones et al. 2012; P. Gratier et al. 2017), chemical abundances (e.g., Y. S. Ting et al. 2012), and dark matter halos (e.g., A. W. C. Wong & J. E. Taylor 2012; Y. Chen et al. 2020). Here, we limit ourselves to a brief description of PCA, mainly via the singular value decomposition and its application to modeling SFHs. In particular, we adopt the PCA algorithm from scikit-learn (F. Pedregosa et al. 2011), as described below.

Consider a sample of $N_{\rm G}$ galaxies and let the SFH of the $i$th galaxy be $f_i(t)$, where $i = 1, 2, \cdots, N_{\rm G}$. We sample the time $t$ in $N_t$ points, $t_j$, with $j = 1, 2, \cdots, N_t$. Thus, the SFH of the $i$th galaxy can be considered as a "vector" in the $N_t$-dimensional hyperspace, $f_i \equiv [f_i(t_1), f_i(t_2), \cdots f_i(t_{N_t})]$, with $f_i(t_j)$ being the projection of the "vector" along the $j$th axis. In practice, we work with the normalized SFH, defined as

$$\hat{f}_i(t_j) = \frac{f_i(t_j)}{\text{mean}(f_i)}, \tag{6}$$

where the mean($f_i$) is calculated by the total stellar mass over the age of the Universe. To simplify the notation, we will omit the hat on $\hat{f}$ and use $f$ to represent the normalized quantity. Thus, the SFH data set of the sample is an $N_{\rm G} \times N_t$ matrix, $A$, with each row representing the normalized SFH of one galaxy in the sample.

Fed with the data of a given training sample, the first step of PCA is to determine an empirical mean vector (of length $N_t$). This mean vector is used to shift the origin of the original coordinate system to a new point represented by the mean vector. In particular, the shape of this mean vector is similar to the cosmic SFR, though not identical, because of the limited training sample of our study. Then, the transformation of the old coordinates into the new orthogonal basis is done by a covariance matrix,

$$C(t_k, t_l) \equiv A^T A = \sum_i f_i(t_k) f_i(t_l), \tag{7}$$

which has a dimension of $N_t \times N_t$. The new coordinates, i.e., the eigenvectors (eigenhistories, in our study), and eigenvalues are found by performing a singular value decomposition over the covariance matrix:

$$C = V \Lambda V^T, \tag{8}$$

where the $k$th column of $V$ is the $k$th eigenhistory, denoted by $e_k$, and $\Lambda$ is the diagonal matrix containing the eigenvalues, $\lambda_k$, that represent the relative importance of each eigenvector. From the PCA construction, the eigenvalues are ranked from high to low, indicating that the first few eigenvectors govern the most dominating features of the training data set. In general, the original data set can be reconstructed using a linear combination of the eigenvectors with corresponding coefficients, $a_k$, and the constant mean vector, i.e.,

$$f_i(t_j) = \sum_{k=1}^{N_e} a_{i,k} e_k(t_j) + \text{mean vector}, \tag{9}$$

where $N_e$ is the total number of eigenvectors used in the reconstruction. It is straightforward to see that $N_e = N_t$ if all eigenhistories are used. Since one of our goals is to use only the most important eigenhistories (instead of all of them) to reduce the dimensionality of the SFH representation, we only select the first few eigenhistories, based on how much variance they can recover from the original data set. For reference, we examine the PCA-based models when using various numbers of eigenhistories in Appendix B.

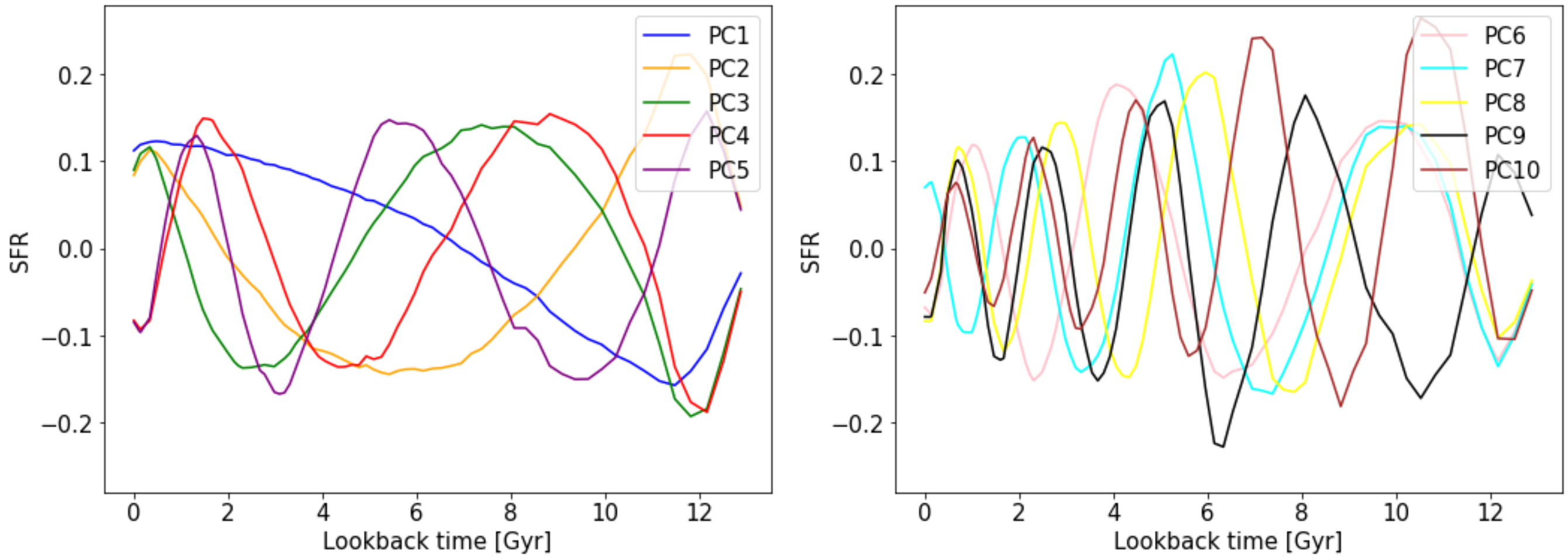

**Figure 2.** The first 10 eigenhistories of the PCA-based models. The left panel shows the lower-order eigenhistories from one to five, while the right panel shows the higher-order eigenhistories from six to 10. The eigenhistories are sampled in the logarithmic lookback timescale.

*3.2.2. The Development of a PCA-based Model for SFHs*

As mentioned above, the original TNG100 data, given in redshift, are preprocessed into cosmic time (and lookback time), given the cosmological model adopted. We develop PCA-based models as follows.

*The PCA model*. Since it is known that galaxy SEDs are lightweighted SFHs, the most recent star formation naturally has a stronger influence on the spectral shape than older stellar populations. We will discuss this effect in more detail later, in Section 4.3. This is also the reason that nonparametric models usually adopt a denser sampling at a later (cosmic) time. We therefore consider a model in which the lookback time, $t_{\rm lb}$, is resampled on a grid of 100 points on a logarithmic scale. In doing so, the most recent SFRs are weighted more importantly relative to the earlier SFRs (see Figure 1). The first 10 eigenhistories obtained are shown in Figure 2. As one may see, the eigenhistories oscillate around SFR = 0.0, more or less as sinusoidal waves. As the eigenhistories go to higher orders, the oscillation becomes more frequent, indicating that lower-order eigenhistories represent the overall trend of an SFH, while the higher orders represent the small-scale details. Furthermore, oscillations are more frequent at smaller $t_{\rm lb}$, especially for higher-order eigenhistories. This arises from the higher weight given to the more recent cosmic time in the PCA model.

We perform a percentage-of-variance-explained (PVE) test on the PCA model. Since the TNG100 training data set is interpolated on a grid of 100 temporal points, there should be 100 independent eigenhistories in total. If all of the eigenhistories are included, the sum of the PVE is 1. In Figure 3, we plot the cumulative PVE (CPVE) from the first eigenhistory to the 15th. Clearly, a small number of the low-order eigenhistories already represent a significant amount of the variance of the data set. Although the CPVE value continues to approach 1 as more eigenhistories of higher orders are included, the growth of the CPVE within the first few eigenhistories is faster than that in higher orders. In Appendix B, we statistically study the effect of using different numbers of eigenhistories in the PCA-based models and find that five eigenhistories provide an ideal balance between model simplicity and accuracy in describing the simulated SFHs, while also matching the corresponding mock spectra. Thus, we use five eigenhistories in the PCA-based models throughout the paper, unless specified otherwise. As shown in Figure 3, this choice can capture ∼65% of the variance in the original TNG100 data.

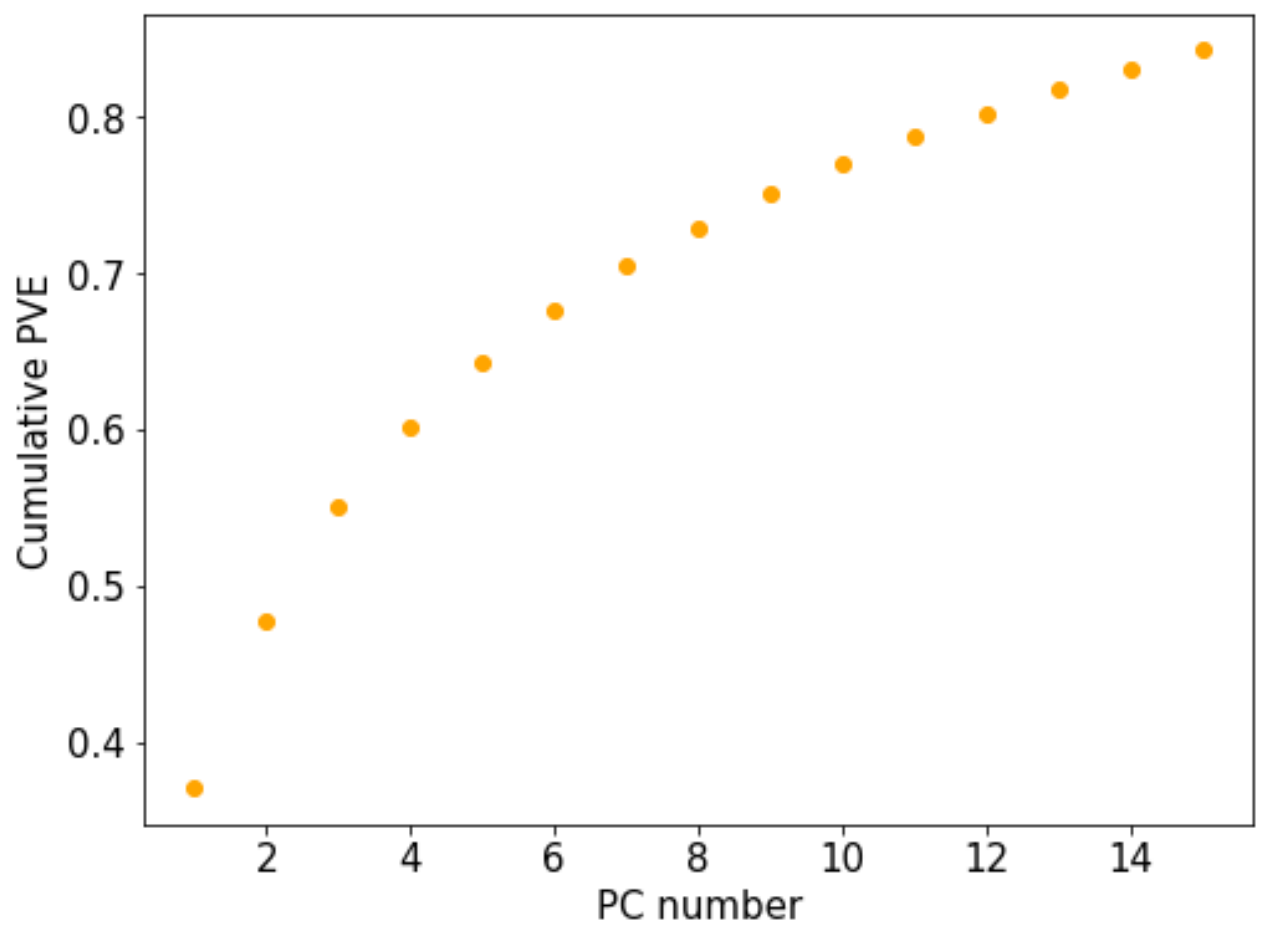

**Figure 3.** CPVE of the PCA model. The first five eigenhistories capture ∼65% of the variances in the original TNG100 data.

*The PCA+step model*. Owing to the fact that each eigenhistory is its own entire history as a function of time, there is a strong entanglement between the old and young stellar populations in the PCA model and thus also in the composite spectrum. Since a younger population is brighter than an old population of the same total mass, particularly when the stellar age is younger than ∼1 Gyr, the presence of a recent burst of star formation can affect the spectrum of a galaxy significantly, even if the total amount of stars formed in the burst is small. This can lead to significant biases in the inferences based on the spectrum. To deal with this problem, we modify the PCA model by including the freedom to tune the SFR in the most recent time interval (a step function). We choose a step that covers 0–0.3 Gyr in lookback time, based on our tests using different step widths in Section 4.3. This step adds an extra degree of freedom to the model. Thus, the PCA +step model has the degree of freedom of the PCA model plus 1, which is the same as the stepwise model described above.

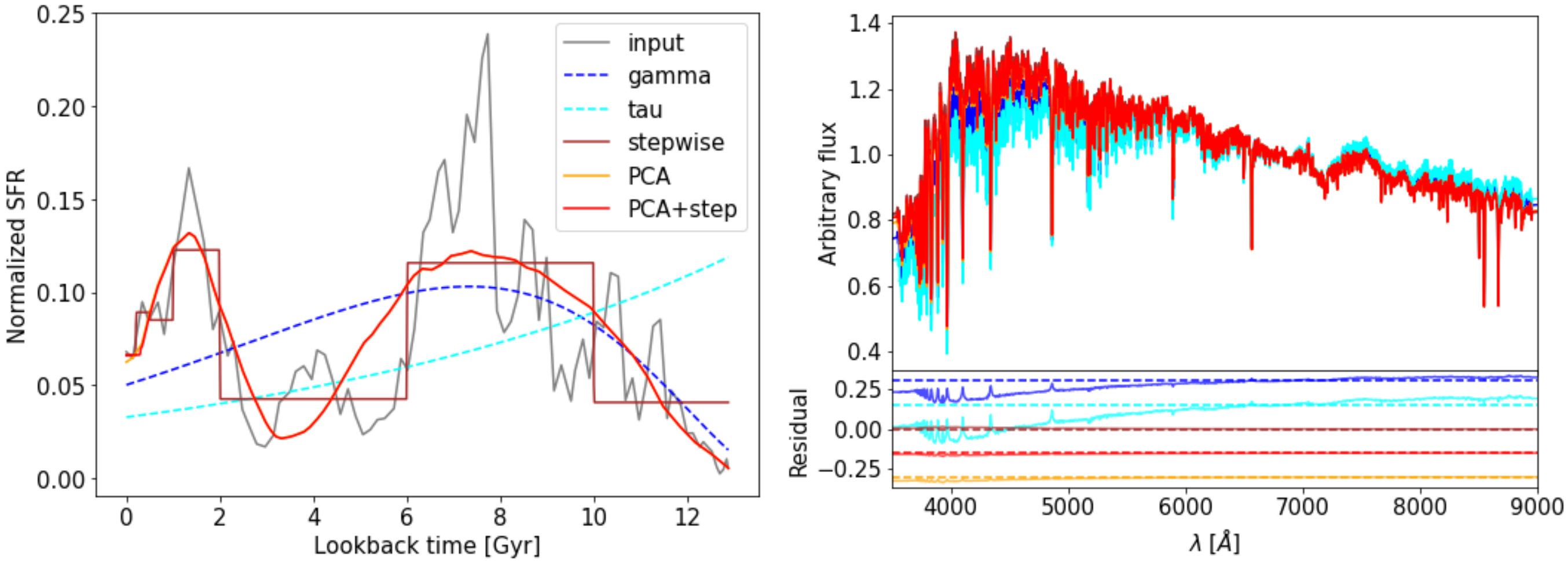

**Figure 4.** Left: an example SFH (gray curve) given by TNG100 and the best fits to it assuming different models. All the SFH curves are normalized to 1 $M_{\odot}$. Right: spectrum generated from the original SFH in comparison to those generated from the best-fit SFH models. The spectra are normalized using the mean flux density between 6500 Å and 7000 Å. Residuals of the model spectra relative to the spectrum of the input SFH are shown at the bottom of the right panel, with the zero-points of the $\Gamma$, $\tau$, and PCA-based models shifted by constants.

## 4. Comparisons between Models

A successful SFH model should be able to reproduce the diversity of galaxy SFHs. Since the SFH of an observed galaxy is usually inferred from its spectrum, and since the observed spectrum is the lightweighted composite of stellar populations at different ages, another criterion is that the model should not lead to biased inferences of the SFH from the spectrum. Because an SFH model is an approximation to the true SFH, the best fit to the SFH itself does not necessarily mean the best match to the spectrum, and vice versa. It is thus essential to test SFH models in both the SFH space and spectral space. This can be done only for simulated galaxies where the SFHs for individual galaxies are available. In this section, we test the representative SFH models described above alongside the new PCA-based models, in both spaces. To this end, we first fit SFHs from TNG100 to different models and examine their goodness of fit. We then compare the spectra generated by each best-fit SFH to the true spectrum generated using the simulated SFH.

### 4.1. Comparisons in SFH Space

We use the full SFHs of individual galaxies as provided by TNG100, and we use a least-squares fit to fit each SFH (in linear scale) with each of the SFH models described earlier. As mentioned before, we only use data at $z \lesssim 6$. The fitting procedure uses the Python package NonLinear Least-squares Minimization and Curve-fitting (LMFIT; M. Newville et al. 2023). Figure 4 shows one example. The left panel shows the best-fit models in comparison to the original input SFH (gray). As seen from this example, the input SFH has two burst periods, with a major and earlier peak around the lookback time $t_{\rm lb} \sim 7$ Gyr and a secondary one around $t_{\rm lb} \sim 1$ Gyr. The $\Gamma$ model (shown by the dashed blue line) reproduces the earlier peak, but the power-law behavior of the model at lower $t_{\rm lb}$ fails to recover the secondary peak. The $\tau$ model (shown by the dashed cyan line) also fails, because it only allows an exponential decay at small $t_{\rm lb}$. Overall, the two parametric models can roughly recover the general trend of the decreasing SFR with time, but they cannot capture fluctuations in the input SFH. The stepwise model, shown by the brown line, performs reasonably well in capturing the shape of the input SFH at low $t_{\rm lb}$, but it only provides a coarse approximation at high $t_{\rm lb}$ because of the large steps used.

In contrast, the PCA-based models successfully recover the overall trend, as well as significant higher-order features, such as the two peaks. The two PCA-based models return quite similar least-squares results, as they only differ in the step covering $t_{\rm lb}$ = 0–0.3 Gyr. Within the past 0.3 Gyr, the predicted SFR of the PCA model follows a decreasing trend, while adding the step function in the PCA+step model results in better agreement with the SFR in the most recent time bin. As we will show in Section 4.3, recent star formation carries a large weight in the predicted spectrum. It is thus important to check the performance of SFH models also in spectral space, as we will do in the next subsection.

Statistically, the success of a model can be quantified by examining its performance on a sample of galaxies covering a variety of SFHs. To this end, we apply the same least-squares fitting to each of the TNG100 galaxies and define the following quantity, to describe the goodness of match between a model SFH and the original SFH:

$$\Delta_{\rm SFH} = \frac{\sigma}{\mu}, \tag{10}$$

where

$$\mu \equiv \frac{1}{N}\sum_{i=1}^{N} {\rm SFR}_{{\rm input},i} \tag{11}$$

is the average SFR over the SFH, and

$$\sigma^2 \equiv \frac{1}{N}\sum_{i=1}^{N} ({\rm SFR}_{{\rm input},i} - {\rm SFR}_{{\rm model},i})^2 \tag{12}$$

is the variance between the input SFH and the best-fit model SFH, with $N$ being the total number of time bins used to

**Table 1**
Percentiles of $\Delta_{SFH}$ and $\Delta_{spec}$ Statistics for TNG100 Central Galaxies in the SFH (First Column) and Spectral (Second Column) Spaces

| Model | $\Delta_{SFH} \times 10$<br>25%–50%–75%–85%–95% | $\Delta_{spec} \times 100$<br>25%–50%–75%–85%–95% | Sample |
|---|---|---|---|
| Γ | 4.2–4.9–6.1–7.0–9.9 | 2.7–5.8–10.5–13.3–18.6 | TNG100 central |
| | (4.8–6.0–8.3–10.8–23.0) | (2.7–6.0–11.6–15.3–22.6) | TNG100 satellite |
| | [2.6–3.1–3.9–4.5–6.4] | [2.3–4.9–8.8–11.3–15.7] | EAGLE central |
| $\tau$ | 6.7–7.9–9.3–10.3–13.3 | 4.7–9.9–18.4–25.7–30.3 | ⋯ |
| | (7.3–8.9–11.6–14.3–26.2) | (8.5–20.1–29.3–30.7–32.4) | ⋯ |
| | [5.6–6.4–8.0–8.8–10.8] | [3.3–7.3–13.9–19.9–27.6] | ⋯ |
| Stepwise | 4.2–5.0–6.3–7.6–11.1 | 0.4–0.8–1.3–1.6–2.4 | ⋯ |
| | (4.9–6.5–9.4–11.9–18.6) | (0.5–0.9–1.4–1.8–3.5) | ⋯ |
| | [3.2–3.9–4.9–5.8–8.3] | [0.3–0.6–1.0–1.6–3.5] | ⋯ |
| PCA | 3.6–4.2–5.3–6.2–8.9 | 1.7–3.5–6.0–7.9–12.0 | ⋯ |
| | (4.1–5.3–7.4–9.6–20.0) | (2.3–5.0–9.4–12.8–20.8) | ⋯ |
| | [2.2–2.7–3.4–3.9–5.8] | [0.8–1.8–3.3–4.4–8.5] | ⋯ |
| PCA+step | 3.5–4.2–5.2–6.2–8.9 | 0.4–0.7–1.1–1.5–2.3 | ⋯ |
| | (4.1–5.2–7.4–9.5–17.7) | (0.5–0.8–1.5–2.0–3.7) | ⋯ |
| | [2.2–2.7–3.4–3.9–5.8] | [0.2–0.4–0.7–0.9–1.3] | ⋯ |

**Note.** The five values from left to right in each entry are the 25%–50%–75%–85%–95% percentiles. The values in parentheses are from TNG100 satellite galaxies, and the values in square brackets are from EAGLE central galaxies.

represent the SFH. We use $N = 87$ for the TNG galaxies, which corresponds to the snapshots returned directly from the TNG simulation.

The left panel of Figure 6 shows the probability density distributions of $\Delta_{SFH}$ for different models. As one can see, the distributions of $\Delta_{SFH}$ for the PCA-based models are similar to each other, peaking at $\Delta_{SFH} \sim 0.38$. This is expected, because these two models differ only by the step covering $t_{lb} = 0$–$0.3$ Gyr. The stepwise model and the Γ model have very similar distributions, both peaking around $\Delta_{SFH} \sim 0.45$. However, since the degree of freedom of the stepwise model is larger than that of the Γ model, the latter seems quite powerful in modeling the SFH. The PCA+step model has the same degree of freedom as the stepwise model, while the PCA model has one degree less. The fact that their $\Delta_{SFH}$ distributions are narrower and have lower mean values suggests that the PCA-based models are more powerful in describing the SFHs of the simulated galaxies used herein. The $\tau$ model performs the worst; its $\Delta_{SFH}$ distribution is much broader, with the largest median $\Delta_{SFH}$ value, and it has a more extended tail toward the large-value end. For reference, we list the values of $\Delta_{SFH}$ ($\times 10$) at several percentiles (25%, 50%, 75%, 85%, and 95%) of the $\Delta_{SFH}$ distribution in the first column of Table 1. We note that the errors on these percentiles are small, as the distribution functions are well sampled by the data.

As another test of the performance of the different SFH models, we calculate the cumulative SFRs of each galaxy in each model and obtain the time, $t_f$, by which a fraction of $f = 30\%$, 50%, and 70% of the stellar mass has formed, respectively. Figure 5 shows the deviation (vertical axis) of $t_f$ between a model prediction and that obtained from the input SFH, with the horizontal axis plotting the corresponding $t_f$ of the input SFH. Results are shown for the different models, as indicated at the top of the plots, and for the three percentiles, as indicated on the right-hand side. The light points are the results for individual galaxies, while the black points connected by thick lines are the median values estimated in six formation time bins, with the error bars indicating the 5%–95% percentile range. A positive (negative) value of the deviation implies that the SFH model predicts a later (earlier) time to form the given fraction of stellar mass than the input SFH.

As seen from Figure 5, the $\tau$ model shows significant systematic deviations from the ground truth in each mass fraction. For galaxies that have a given fraction of their stars formed early (late) according to the input SFH, the $\tau$ model tends to predict a later (earlier) formation time. The performance of the Γ model is much better, even though it has the same degrees of freedom as the $\tau$ model. The median of each time bin given by the Γ model is quite close to zero (indicated by the gray dashed line), with error bars much smaller than those given by the $\tau$ model. We notice that for the formation time of the 70% mass, the predicted formation time tends to be slightly later than the input values for times earlier than $\sim$5 Gyr, indicating that the Γ model does not perform ideally for galaxies that have a significant amount of old stars. For the stepwise model, the overall deviation from the ground truth is small. For all of the mass formation times, this model predicts a slightly earlier formation time than the input values. The two PCA-based models have similar performances, with the predicted medians all close to zero. The scatters represented by the error bars are also quite small, typically about 0.1–0.5 Gyr, and independent of the formation time. All in all, the results shown in Figure 5 demonstrate that the PCA-based models can provide an accurate representation of the simulated SFHs.

### *4.2. Comparisons in Spectral Space*

Since the observed spectrum of a galaxy is its SFH weighted by the light of stellar populations of different ages, we need to check if the predicted spectrum of a best-fit SFH model can also match the spectrum expected from the original input SFH. To test this, we generate mock spectra using the best-fit SFHs and compare them to the spectra generated using the input SFHs. To isolate the effects produced by the SFH, we keep all other factors (IMF, dust, and metallicity, etc.) the same. All the spectra are generated for the full wavelength range of 3500–9000 Å and normalized using the mean flux between 6500 and 7000 Å. An example, generated using the SFHs shown in the left panel, is shown in the right panel of Figure 4. From the residual plot, we see that the spectra generated by both the $\tau$ and $\Gamma$ models significantly underestimate the input spectrum at the blue end, clearly because these two models miss the secondary peak in the SFH at low $t_{\rm lb}$. The stepwise model and the two PCA-based models have similar performances in matching the input spectrum, showing no significant offset from zero in the residual. These results demonstrate that for some cases, such as the one shown here, the performance of a model in spectral space can be quite different from that in SFH space. The difference in the spectra between the PCA model and PCA+step model is almost entirely produced by the small difference of the SFH in the most recent time bin, $t_{\rm lb}$ = 0–0.3 Gyr, indicating that it is important to take into account recent star formation in an SFH model to obtain unbiased inferences of the SFH from the synthesized spectrum.

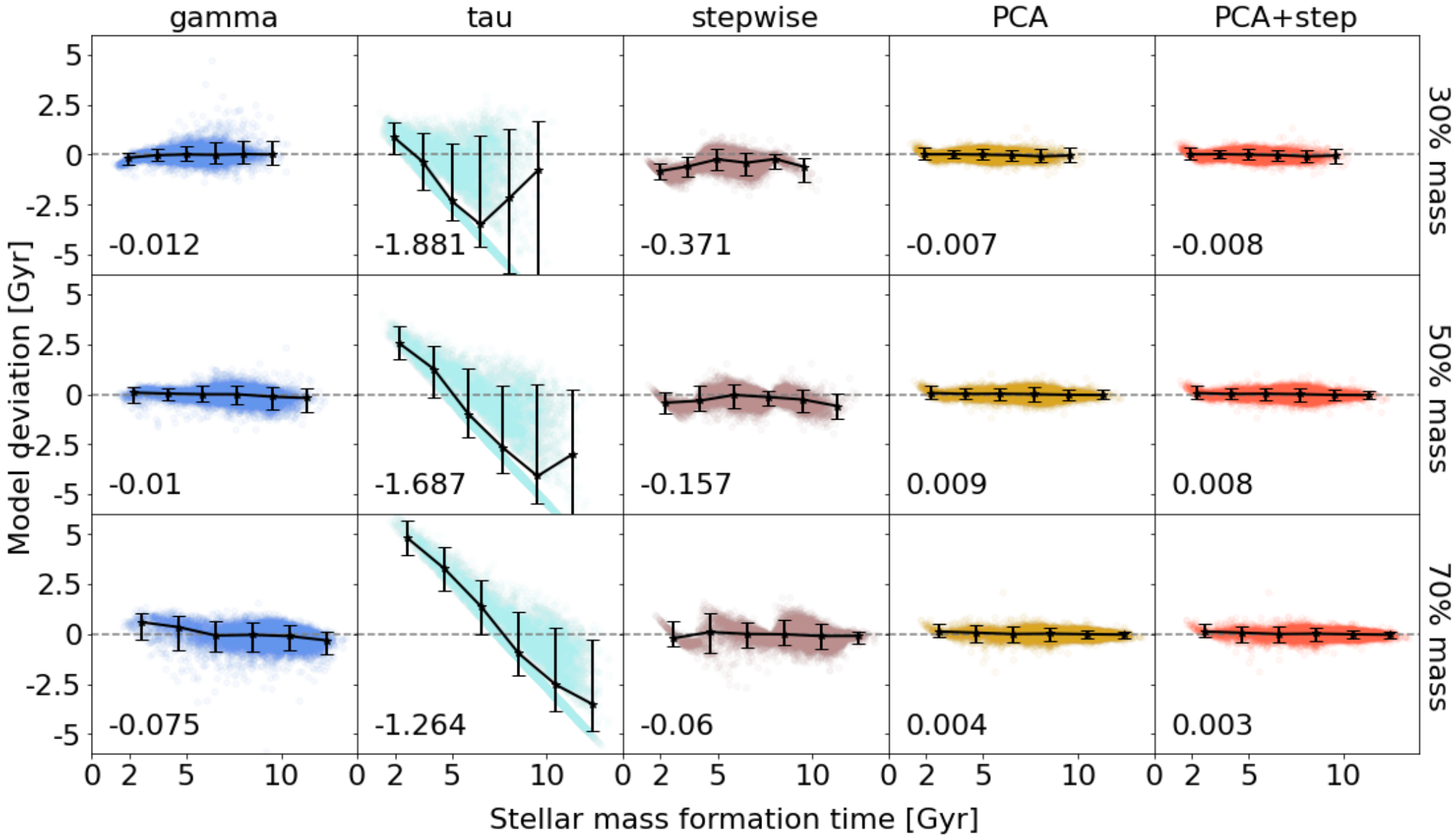

**Figure 5.** Comparing fractions of the mass formation time from each SFH model to the ground truth (marked by the gray dashed line). The *x*-axis marks the mass percentage formation time, and the *y*-axis is the time deviation between the model prediction and that obtained from the input SFH. Each SFH model is indicated by the column title, and each row represents 30%, 50%, and 70% of the total stellar mass formed. In each panel, the light colored dots represent the individual simulated galaxies in the sample. We bin the real formation time in six time bins and mark the median deviation using the black dots. The error bars indicate the 5% and 95% percentiles from the bins. Furthermore, the median value of the model's deviation from the entire data set is labeled at the lower left corner of each panel.

We calculate the $\Delta_{\rm spec}$ value of each model spectrum relative to the input mock spectrum for full spectral coverage using the same equation as shown in Equation (10), but for the flux density instead of the SFR, and obtain the probability distribution of $\Delta_{\rm spec}$ predicted by each of the SFH models for the TNG sample. The percentiles of the distributions are summarized in the second column of Table 1 for each model. It is seen that both the $\tau$ and $\Gamma$ models have the largest percentile values, indicating the poor performances of these two models in spectral space. The percentile values of the PCA model are about four to six times as large as those of the PCA+step model, indicating again the importance of correctly modeling the recent star formation in an SFH model. The stepwise model ranks next to the PCA+step model, with similar statistics in $\Delta_{\rm spec}$, although its performance in SFH space is worse.

The right panel of Figure 6 shows the full distribution of $\Delta_{\rm spec}$ for each model. The distribution obtained from the $\tau$ model extends to very large values, as seen from the slowly growing cumulative distribution. The $\Gamma$ model performs slightly better than the $\tau$ model, but again with a large $\Delta_{\rm spec}$ at the peak and a slowly growing cumulative distribution. These results suggest that these parametric models cannot reproduce the spectra of many simulated galaxies. The stepwise model has much better statistics, as can be seen from the higher probability at the lower $\Delta_{\rm spec}$ end, as well as the relatively fast growth of the cumulative distribution. Roughly 50% of the galaxies in our sample have $\Delta_{\rm spec}$ smaller than ∼0.008 in the stepwise model, in comparison to ∼0.1 for the $\tau$ model and ∼0.06 for the $\Gamma$ model. For PCA-based models, we can see the significant improvement from the PCA model to the PCA+step model, with the $\Delta_{\rm spec}$ distribution shifting from large to small values and the cumulative distribution becoming successively steeper. The PCA+step model returns the best statistics among all models: the median $\Delta_{\rm spec}$ is ∼0.007, slightly smaller than that of the stepwise model, ∼0.008. And more than 55% of the galaxies have their $\Delta_{\rm spec}$ values lower than 0.008. Overall, the stepwise model and PCA+step model perform the best in spectral space, although the stepwise model is less accurate in SFH space (see Section 4.1).

In addition, we test the $\Delta_{\rm spec}$ behavior at various wavelength coverages, e.g., 4MOST blue and green bands (R. S. de Jong et al. 2019), Prime Focus Spectrograph blue and red bands (M. Takada et al. 2014), and various narrow bands, and find that the stepwise and PCA+step models always result in the best statistical behavior compared to the other SFH models, with more significant improvements in the bluer bands. This is consistent with what is shown in the right panel of Figure 4, where the spectral residual shows the strongest wavelength dependency at the blue end. This also indicates a strong influence on the galaxy spectrum from the youngest stellar population, in agreement with Section 4.3.

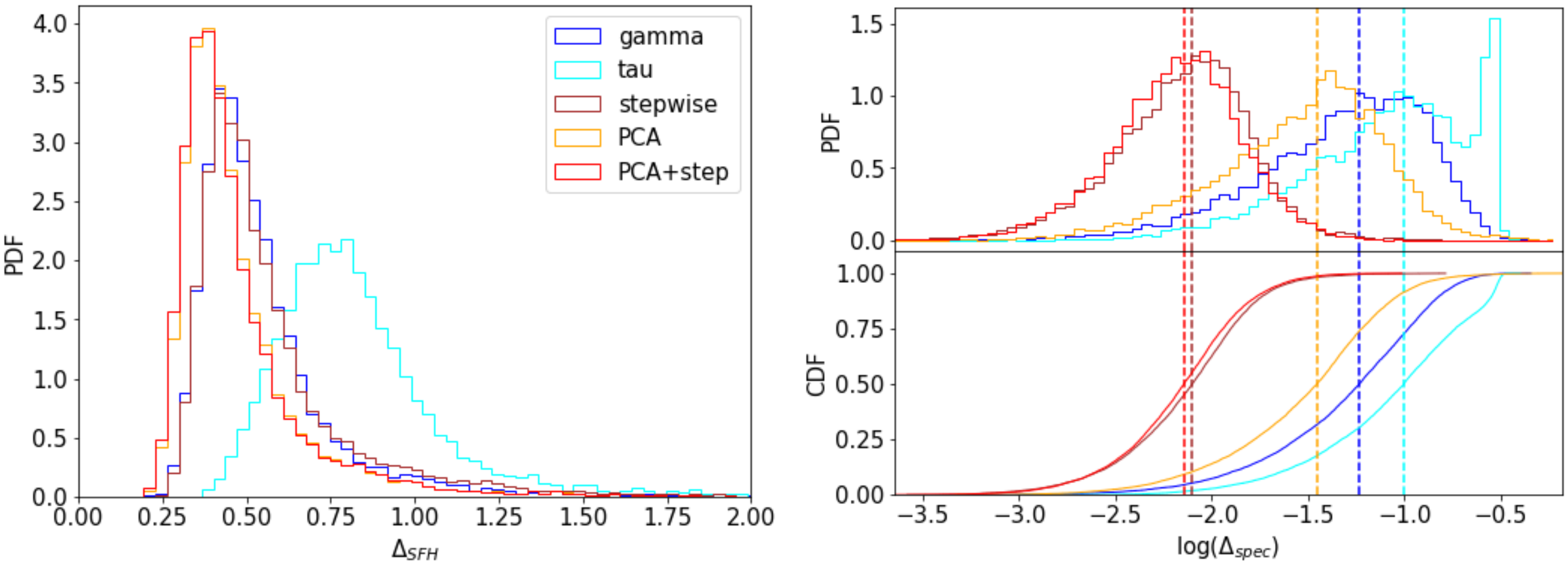

**Figure 6.** Left: probability density distribution of $\Delta_{\rm SFH}$. The PCA-based models share similar distributions, with a peak at $\Delta_{\rm SFH} \sim 0.38$, the smallest compared to all the other models. The $\Gamma$ model and stepwise model share similar distributions, both peaking at $\Delta_{\rm SFH} \sim 0.45$, while the $\tau$ model has the largest peak value of ~0.75. Right: the upper panel is the probability density distribution in the log($\Delta_{\rm spec}$) bin, and the lower panel is the corresponding cumulative distribution. The dashed lines mark the median $\Delta_{\rm spec}$ value of each model. The models in the right panel use the same color codes as in the left panel.

Comparing the statistics in $\Delta_{\rm SFH}$ and $\Delta_{\rm spec}$, one sees that the median values in SFH space are generally an order of magnitude larger than those in spectral space, as shown in Table 1. This demonstrates that the differences in SFH space are significantly smeared out in spectral space. Such a "smearing effect" makes it difficult to constrain the SFH of a galaxy from its spectrum and indicates that it is crucial to design SFH models carefully, to obtain unbiased inferences of SFH from synthesized spectra. As pointed out above, this requires an SFH model that performs consistently in both the SFH and spectral spaces. Based on the results presented above, we find that the PCA+step model performs consistently in both spaces. In what follows, we perform further tests of this model in comparison with traditional models.

### 4.3. Impacts of Most Recent Star Formation

As shown in Figure 4, some of the best-fit SFHs produce spectra that are quite different from those of other SFH models. Since all other factors are kept the same, except the shape of the SFH curve, this raises the questions (i) what part of the SFH causes this discrepancy, and (ii) how significant can the effect be? For the example shown in Figure 4, the most noticeable difference from the best-fit SFHs is the most recent SFR, where the two parametric models have noticeably lower SFRs compared to the stepwise model and the PCA-based models. The dramatically different behavior between the PCA model and the PCA+step model in spectral space, shown in Figure 6, reveals the same problem. Since the two PCA-based models only differ by the step function at $t_{\rm lb} = 0$–$0.3$ Gyr, the difference indicates the importance of including the freedom to account for the most recent star formation. The choice of the step width, 0–0.3 Gyr, is not justified a priori. Here, we test the influence of different step widths, specifically from 0.1 Gyr to 0.5 Gyr.

We repeat the same $\Delta_{\rm SFH}$ and $\Delta_{\rm spec}$ tests that we have done in previous sections. As shown in the left panel of Figure 7, the distribution of $\Delta_{\rm SFH}$ is almost independent of the step width, as expected from the fact that the amount of star formation in the most recent time step is small in comparison to the entire SFH. Thus, from the SFH perspective, all of these step widths are equally valid. However, when the best-fit SFHs are used to generate model spectra, it seems that a width of 0.3 Gyr is preferred, as shown in the right panel of Figure 7. The next best width is 0.2 Gyr, then 0.1 and 0.4 Gyr, for which the distributions of $\Delta_{\rm spec}$ have similar median values, although slightly larger than that of the best choice. As the step width increases to 0.5 Gyr, the peak of the distribution is shifted toward larger $\Delta_{\rm spec}$, and the median is about two to three times as large as that of the best choice. These test results show that the choice of 0.3 Gyr for the step width is close to optimal.

### 4.4. Effects of Adding Noise

The input mock spectrum shown in Figure 6 contains no noise. However, measured signals always come with noise in real observations. Therefore, we test the acceptability of the selected SFH models at various SNRs. To this end, we add Gaussian noise with zero mean and a standard deviation $\sigma_{\rm noise}$ to the input mock spectra, thus generating spectra with $\mathrm{SNR} = 1/\sigma_{\rm noise}$ per angstrom. We vary the SNR from 2 to 40 (and 100 for the comparison between the stepwise and PCA +step models) and calculate the reduced chi-square in the predicted spectrum for each SFH model relative to the ground truth. The $\chi^2_\nu$ is calculated using

$$\chi^2_\nu(\mathrm{SNR}) = \frac{1}{N-1}\sum_i^N \frac{(f_{{\rm input},i} - f_{{\rm model},i})^2}{\sigma^2_{\rm noise}} \tag{13}$$

and is shown in the top panel of Figure 8. We plot the 50% percentiles using solid lines and the 25% and 75% percentiles using error bars. For reference, we also show the spectrum of the galaxy shown in Figure 4 with noise at SNR = 2 and SNR = 40 in comparison to the spectra given by different SFH models.

Using the best-fit SFH, the $\chi^2_\nu$ of the $\Gamma$ and $\tau$ models decreases rapidly with the SNR. This can be understood by comparing the spectral plots in Figures 4 and 8. When the predicted spectra are compared with the noise-free input spectrum in Figure 4, the best-fit SFHs of the $\Gamma$ and $\tau$ models perform poorly in spectral space. At SNR = 40, the discrepancies of these models at the short-wavelength end with the zero-points defined by the ground truth are still very significant. However, as the input mock spectrum becomes highly noisy, e.g., at the SNR = 2 level, all of the SFH models are "acceptable" in spectral space. As one can see from the top panel, showing $\chi_\nu^2$ as a function of the SNR, a moderate SNR is required to reject the $\tau$ and $\Gamma$ models, but a very high SNR is needed to reject the stepwise model and the PCA+step model. This demonstrates that the inference of the SFH from a noisy spectrum can be biased by using a restrictive model for the SFH, such as the parametric models.

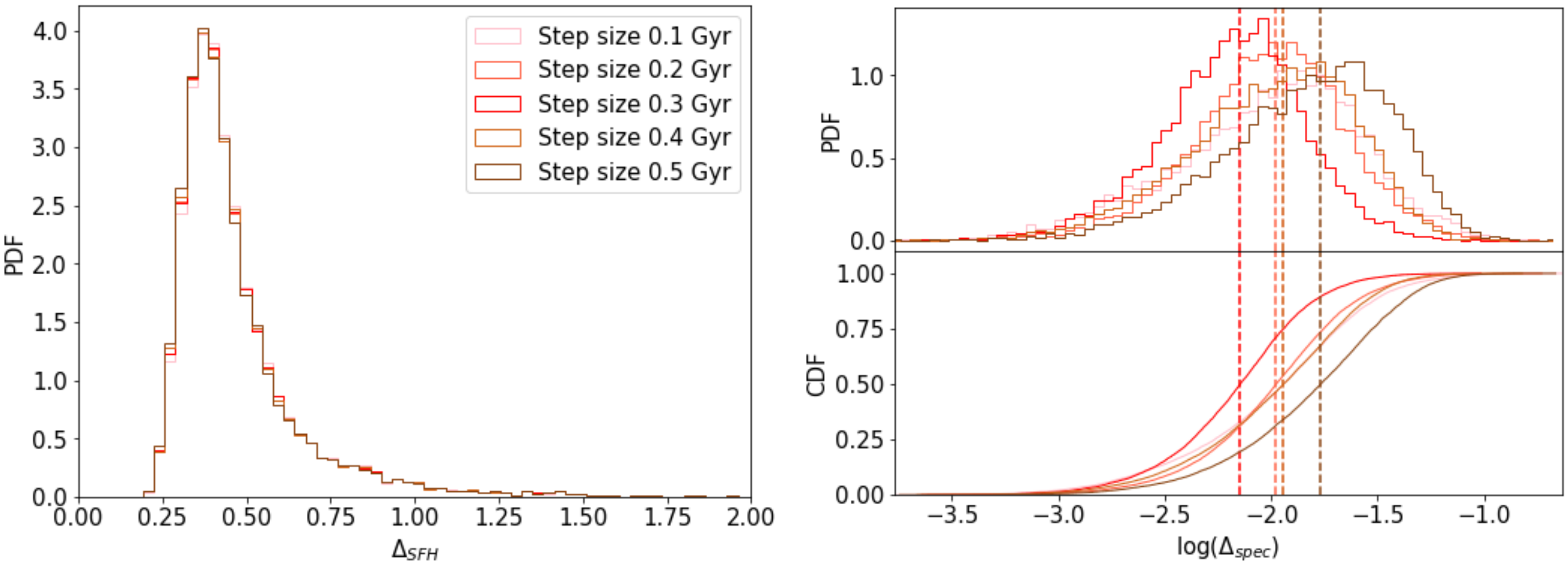

**Figure 7.** Probability density distributions of $\Delta$ for the PCA+step model with various step sizes (0.1–0.5 Gyr) in SFH space (left panel) and spectral space (right panel). The 0.3 Gyr step size model (plotted in red) is the PCA+step model shown in the previous plots.

It is interesting to see that the PCA model, which is flexible enough in describing the overall SFH and can accurately predict the formation times of fixed fractions of stellar mass (Figure 5), does not perform as well as the PCA+step model in spectral space. As shown in Figure 8, the growth rate with increasing SNR of the PCA model is the next fastest besides the two parametric models. About 50% of the model spectra can be rejected (using $\chi_\nu^2 > 1$ to indicate a significant discrepancy) at SNR = 3–5, and 75% can be rejected at SNR = 5–10. In contrast, the PCA+step model makes a large improvement in spectral space: it has 50% of the spectra rejected at SNR = 5–10 and 75% at SNR = 40.

Comparing the stepwise model and the PCA+step model, one sees that the PCA+step model is slightly favored starting from SNR = 20 to higher SNRs. Overall, these two models have similar performances in spectral space, even in the presence of noise.

## 5. The Applicability of PCA-based Models

So far, we have demonstrated the validity of the PCA-based models using the same data set as used for the model training (i.e., to obtain the eigenhistories). Yet, it is important to test the validity of the models using different populations of galaxies. In this section, we first test the models using the central galaxies in three stellar mass bins in the TNG100 simulation (Section 5.1). We then test in Section 5.2 the validity of the models on TNG100 satellite galaxies that are not used for training the PCA-based models. Finally, in Section 5.3, we test models using central galaxies selected from the EAGLE simulation.

### 5.1. Tests on Different Stellar Mass Bins

To this end, we divide the selected central galaxies from the TNG100 sample into three mass bins—low-mass ($M_* < 10^{10}h^{-1}M_\odot$), intermediate-mass ($10^{10}h^{-1}M_\odot \leqslant M_* < 5 \times 10^{10}h^{-1}M_\odot$), and high-mass ($M_* > 5 \times 10^{10}h^{-1}M_\odot$)—and run the $\Delta_{\rm SFH}$ and $\Delta_{\rm spsc}$ tests for each subsample. The results for SFH space and spectral space are shown in the left and right panels of Figure 9, respectively.

In SFH space, the $\tau$ model performs the worst in all mass bins. The performances of the $\Gamma$ and stepwise models are comparable in each mass bin, although the latter performs slightly poorer for high-mass and intermediate-mass galaxies. This may be a result of the original design of the stepwise model, since more massive galaxies tend to form their stars earlier, while the stepwise model puts more weight on later star formation in a galaxy. Overall, the two PCA-based models have the lowest median of $\Delta_{\rm SFH}$ in each mass bin. There is a common trend in the PCA-based models for different mass bins: the median of $\Delta_{\rm SFH}$ is the smallest for the low-mass sample and the largest for the high-mass sample. This is expected from the design of the PCA-based models: more than 70% of the TNG100 galaxies are categorized as low-mass galaxies, while less than 5% of the galaxies are categorized as high-mass ones. The training sample we used to obtain the eigenhistories is therefore dominated by low-mass galaxies and thus features in their SFHs are better represented. However, such a trend is also seen in the traditional models that are not influenced by the training process. This indicates that low-mass galaxies in the TNG simulation on average have simpler SFHs than higher-mass galaxies. It is also interesting to note that the performance ranking of a particular model does not change significantly between the three mass samples, indicating that the diversity in the SFHs depends systematically on galaxy mass. The preserved ranking order of the tested SFH models provides evidence that the good performance of PCA-based models in SFH space is mass-independent and flexible.

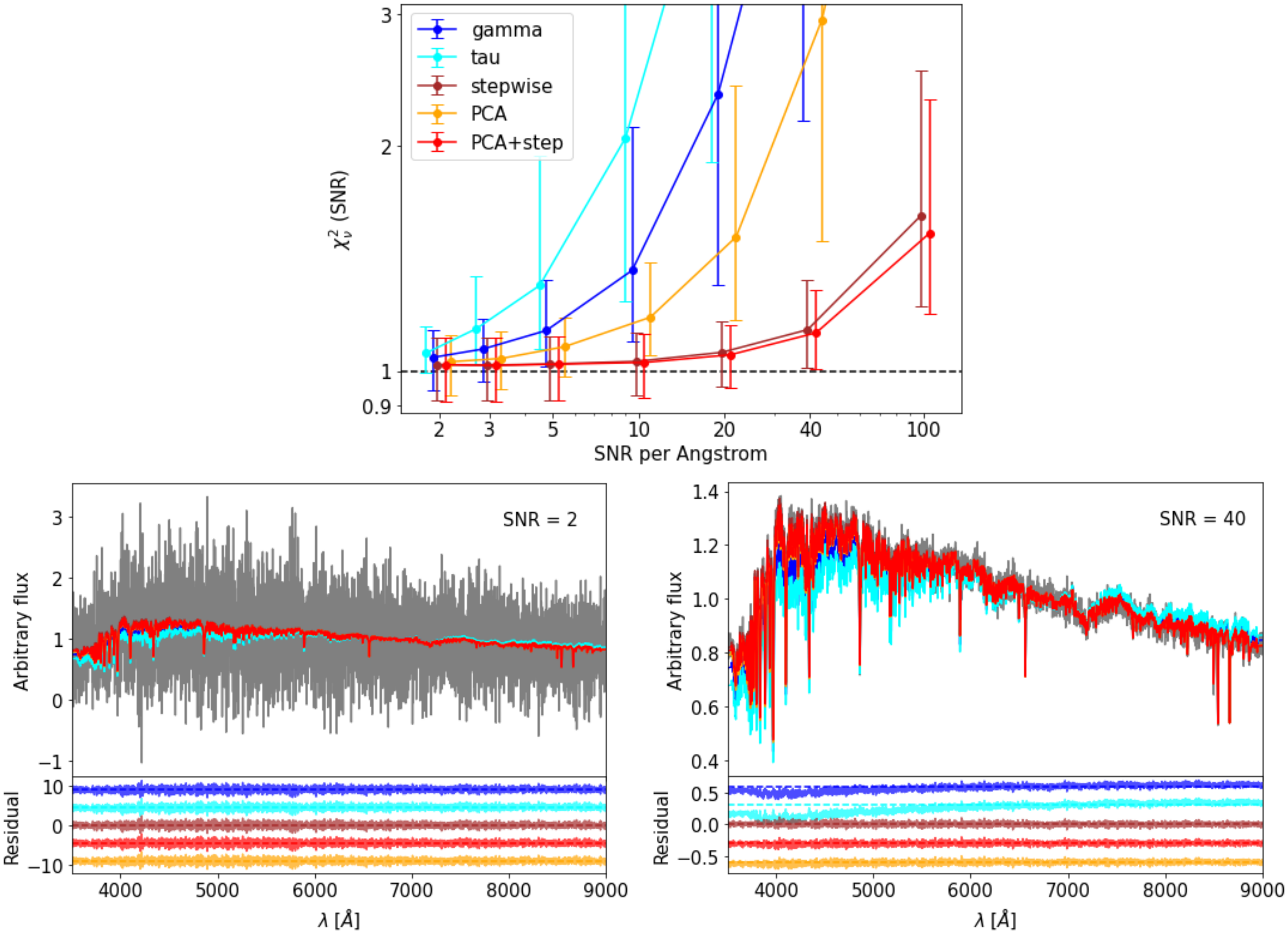

**Figure 8.** Top: the $\chi^2_\nu$ statistics of the TNG100 galaxy spectral comparison at various SNRs (per wavelength). The spectra are generated using the best-fit SFHs of the $\Gamma$ model (blue), the $\tau$ model (cyan), the stepwise model (brown), the PCA model (orange), and the PCA+step model (red). The solid lines mark the 50% percentile of each model, while the error bars mark the 25% and 75% percentiles. We shift each model curve horizontally so that the percentile markers can be clearly seen at each SNR. Bottom: the same galaxy spectral comparison as shown in Figure 4 at SNR = 2 and SNR = 40 levels. The zero-points of each residual are shifted up or down by constants.

The $\Delta_{\rm spec}$ statistics are quite different from those in SFH space. For the $\Gamma$ model, the smallest median value of $\Delta_{\rm spec}$ comes from the high-mass sample, while the largest median comes from low-mass galaxies. In contrast, for the $\tau$ model, the largest median $\Delta_{\rm spec}$ comes from high-mass galaxies. Overall, the median $\Delta_{\rm spec}$ values of the two parametric models are the largest among all models. Remarkably, for the stepwise model, the median $\Delta_{\rm spec}$ values of different mass bins are all similar to each other, suggesting that this model works equally well for different galaxy masses in spectral space. This indicates again that the early star formation, which may depend systematically on stellar mass, has little influence on the spectrum. Furthermore, the stepwise model reproduces the input mock spectra quite well, as indicated by the small values of $\Delta_{\rm spec}$ for all mass bins. The improvement of the PCA+step model over the PCA model is significant and similar for all mass bins in spectral space, although the statistics of the two models are very similar in SFH space. This suggests that the most recent star formation (within ∼0.3 Gyr) plays an important role in spectral space for galaxies of different masses. Finally, the median $\Delta_{\rm spec}$ values predicted by the PCA+step model are slightly smaller than those predicted by the stepwise model in all mass bins, indicating that the improvement made by PCA+step is independent of galaxy mass.

### 5.2. Test on Satellite Galaxies

Satellite galaxies are typically less massive and more affected by environmental effects, such as ram pressure stripping and tidal interactions, resulting in their being gas-poor and quenched (e.g., S. M. Weinmann et al. 2009; Y.-j. Peng et al. 2012; T. Buck et al. 2019; Y. Ding et al. 2024). As a result, such environmental processes that are not significant for central galaxies may have systematic effects on the shapes of satellite SFHs. It is thus important to use satellite galaxies to test the validity of the PCA-based models that are trained by central galaxies. To this end, we select from TNG100 satellite galaxies at $z = 0$ with stellar masses above $1.0 \times 10^9 h^{-1} M_\odot$. This gives a total of 6852 satellites.

The same $\Delta$ analyses as described in Sections 4.1 and 4.2 are carried out for the TNG100 satellite galaxies, and the results are shown by the blue lines in Figure 10 and Table 1. Similar to the previous tests based on TNG100 central galaxies, the performances of the $\Gamma$ and stepwise models in SFH space are similar: the $\tau$ model is the worst, while the two

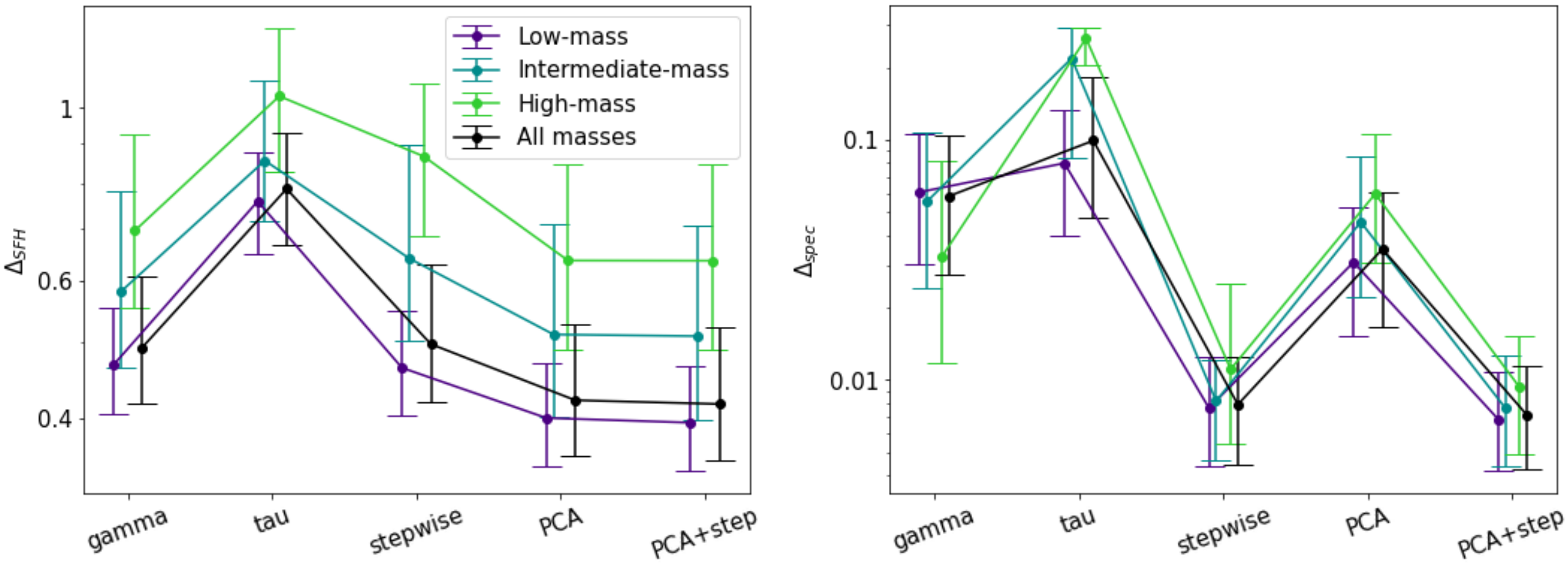

**Figure 9.** The median $\Delta$ values of each SFH model in the low-mass ($M_* < 10^{10}h^{-1}M_\odot$), intermediate-mass ($10^{10}h^{-1}M_\odot \leqslant M_* < 5 \times 10^{10}h^{-1}M_\odot$), and high-mass ($M_* > 5 \times 10^{10}h^{-1}M_\odot$) bins in SFH space (left) and in spectral space (right). The $\Delta$ statistic for all mass galaxies is also shown in the figures as the black curve for reference. The upper and lower error bars mark the 25% and 75% percentiles of the distributions.

PCA-based models perform equally and the best. As one can see from Table 1, the percentile values of $\Delta_{\rm SFH}$ predicted by the PCA-based models for satellite galaxies are all higher than for central galaxies. This can be understood, since the training data come from TNG100 central galaxies only. However, the same pattern is also seen in the traditional SFH models, indicating that environmental effects on satellite galaxies make their SFHs more diverse. But again, the PCA-based models statistically outperform other models, regardless of the different dominating physics in these two types of galaxies in TNG simulations.

In spectral space, the $\Delta_{\rm spec}$ values plotted in the right panel of Figure 10 and listed in Table 1 show that the performance of the $\tau$ model becomes significantly worse for satellite galaxies, while for all other models, the performance for satellites is only slightly worse than for centrals. This is in contrast to the performance in SFH space, where all models become apparently worse for satellite galaxies. This indicates again the nonlinear correlation and degeneracy of SFHs in spectral space. All in all, the PCA-based modeling, typically the PCA +step model, is still preferred by satellite galaxies in the TNG100 simulation.

### 5.3. Test on EAGLE Galaxies

We further test the robustness and flexibility of the new PCA-based models using the EAGLE simulation (J. Schaye et al. 2014; R. A. Crain et al. 2015), which is another cosmological hydrodynamic simulation running on a modified version of the $N$-body Tree-PM smoothed particle hydrodynamics code (V. Springel 2005). We use L100N1504, which simulates galaxy evolution from $z = 20$ to $z = 0$, in a cubic box of side length $L_{\rm box} \sim 100\,{\rm cMpc}$, with a mass resolution $m_{\rm gas} \sim 1.81 \times 10^6 M_\odot$. The cosmology model adopted in EAGLE is from the Planck 2013 results (Planck Collaboration et al. 2014).

Out of the 10,860 galaxies with stellar mass $M_* \geqslant 1 \times 10^9 h^{-1}M_\odot$ at $z = 0$, we select only central galaxies, giving a total of 6160 galaxies. Again, we only use snapshots at $z \lesssim 6$. The total number of snapshots is 23 in the EAGLE simulation, much smaller than the 87 used for TNG100. The lower temporal resolution acts like a smoothing effect on the resulting SFHs, where small-scale features are smeared out. Furthermore, with various different simulation subscriptions, e.g., stellar feedback (TNG: kinetic wind model; EAGLE: stochastic thermal feedback), AGN feedback (TNG: kinetic and radio modes; EAGLE: single-mode thermal feedback), and black hole accretion models, etc. (R. A. Crain et al. 2015; R. Weinberger et al. 2016), the EAGLE SFHs are found to be intrinsically smoother (in lower-mass galaxies) and have extended star formation activities (in higher-mass galaxies; e.g., K. G. Iyer et al. 2020).

The low temporal resolution in the EAGLE SFHs is particularly acute for the stepwise model, because of the missing data in the past 1.35 Gyr. Since the outshining effect is of particular importance during this period, the stepwise model may be inaccurately disfavored in spectral space. Thus, to make a more faithful comparison between the two simulations —specifically, to compensate for the SFH differences due to the different temporal resolutions—we interpolate the EAGLE data onto the TNG snapshots. Furthermore, to mimic the short timescale fluctuations that are not captured by EAGLE, due to its lower temporal resolution, we add random fluctuations using a log-normal distribution obtained from TNG100 central galaxies at individual snapshots. Specifically, the log-normal distributions at different snapshots are obtained by modeling the differences between the original SFHs of the individual TNG100 central galaxies with the corresponding smoothed SFHs, obtained from interpolating the original TNG100 SFHs onto the EAGLE sampling and then interpolating back onto the TNG100 sampling. Such a modification does not contain any TNG-based physics, but rather statistical noise-like fluctuations that may make the SFH-space comparison more comparable. Unfortunately, we cannot further modify the EAGLE SFHs to overcome the differences caused by the different subgrid physics implemented in the simulation. Yet, testing the TNG-trained PCA-based models against the EAGLE SFHs serves as a validation test of whether our models can successfully recover the diversity of different SFHs.

Using the same methodology as for the TNG100 galaxies, we study the $\Delta$ statistics in both the SFH and spectral spaces. The results are shown in both Figure 10 and Table 1. Overall, the performances of the different models for EAGLE galaxies are similar to those for TNG100. For a given model, the value of $\Delta_{SFH}$ for EAGLE tends to be smaller than that for TNG100. This is because the original time resolution of the EAGLE snapshots is lower than that of TNG100, so the SFHs of the EAGLE galaxies are smoothed relative to those of the TNG galaxies when we interpolate the EAGLE data onto the TNG resolution. The short timescale fluctuations we added to the interpolated EAGLE SFHs are assumed to be random, which may underestimate $\Delta_{SFH}$. We also test the EAGLE population with no fluctuations and much higher fluctuations, by adjusting the standard deviations of the log-normal distributions. The ranking order of the SFH models remains the same in all cases, except when the "fluctuations" become unrealistically dominating. In the most extreme regime, all the SFH models become highly comparable in SFH space, because none of them are designed to capture small-scale dominating features. In spectral space, the performances of the different models are very similar between EAGLE and TNG100. In particular, the two PCA-based models perform well for EAGLE galaxies, giving more support to the validity of the TNG-trained eigenhistories in modeling the SFHs of galaxies.

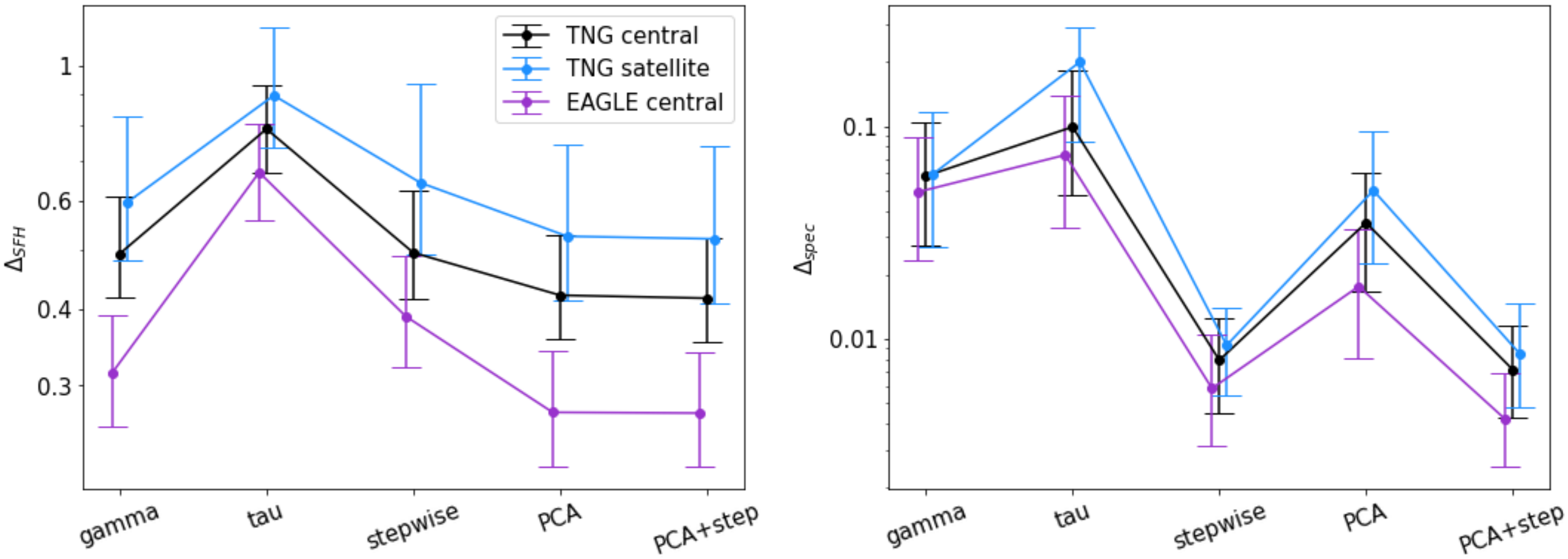

**Figure 10.** $\Delta$ statistics in comparing different simulated galaxy populations, i.e., TNG100 centrals, TNG100 satellites, and EAGLE centrals, in SFH (left) and spectral (right) spaces. The dots mark the 50% percentile values, and the error bars mark the 25% and 75% percentile values.

## 6. Summary and Discussion

In this paper, we examine the statistical behavior of some commonly used SFH models (the $\Gamma$, $\tau$, and stepwise models) in fitting simulated SFHs (SFH space) and compare their synthesized spectra to the "ground-truth" (input) spectra (spectral space). Meanwhile, we develop a new generic SFH model, using PCA (the PCA-based model), trained by central galaxies selected from the state-of-the-art cosmological hydrodynamic simulation IllustrisTNG. The key results are summarized as follows.

For the development of the PCA-based models, we obtain eigenhistories (the PCs). According to the PVE test (Figure 3), the PCA-based models can explain ~65% of the data variance by using the first five eigenhistories. We suggest that this choice is optimal in order to balance model simplicity and accuracy (see Appendix B).

To test the flexibility of the SFH models, we use a least-squares fitting methodology to fit the SFH models to the input SFHs given by TNG100. Statistically, we define a goodness-of-fit metric $\Delta_{SFH}$, to represent the overall deviation of the best-fit models from the input SFHs. We find that the PCA-based models fit the simulated SFHs well, followed by the $\Gamma$ and stepwise models, with the $\tau$ model being the least flexible (Figure 6). The same conclusion is also reached by comparing the simulated and predicted formation times by which a fixed fraction of the total stellar mass is formed in a galaxy (Figure 5).

Using the best-fit SFHs and the input SFHs, we generate synthesized mock spectra and compare between them. We again use a goodness-of-fit metric, $\Delta_{spec}$, to characterize the model performances in a statistical manner. We find that the stepwise and PCA+step models match the input mock spectra well, while the other models do not have the flexibility to cover the diversity, in spectral space (Figure 6). We find that the most recent star formation activity significantly impacts the spectral shape, which can be modeled well by including in the SFH a step in age between 0 and 0.3 Gyr (Figure 7). We also examine the effects of adding random Gaussian noise to the input mock spectra to mimic observational data. We find that high-SNR spectra (at least $SNR > 20$) are required to distinguish the stepwise and PCA+step models that perform the best in spectral space (Figure 8).

We test different SFH models by applying them to different galaxy populations (Figure 10). We find similar model behaviors in both the SFH and spectral spaces for different populations of galaxies, indicating that our test results are valid without depending on the galaxy population in question.

Comparing the model behaviors in SFH and spectral spaces, we notice that the performance of a given model in these two spaces may differ significantly. This demonstrates the nonlinear connection between an SFH and the associated spectrum, suggesting that the performance of a model should be tested in both spaces and that a careful choice of the SFH model is needed to infer the SFH from the observed spectrum through SSM.

It has long been noticed that SFH models can be prior-sensitive, thus various studies investigate the effects of imposing different priors in SFH construction/reconstruction. For instance, J. Leja et al. (2019) comprehensively explore the effect of priors on the shapes of posterior distributions of nonparametric models. Furthermore, B. Wang et al. (2023) extend the study with an empirical constraint, the cosmic SFR density, to encourage rising SFRs in the early Universe and declining SFRs in the later Universe. K. G. Iyer et al. (2019) impose a Gaussian process based on a dense basis SFH reconstruction method (K. Iyer & E. Gawiser 2017) to encourage smooth SFH shapes. Furthermore, N. Caplar & S. Tacchella (2019) model SFHs as stochastic processes with fluctuations around the main sequence and statistically study the variation of SFHs.

The PCA-based models offer a flexible data-driven framework for representing complex SFHs, efficiently capturing the dominant modes of variability found in cosmological simulations such as TNG. This compact basis accommodates a wide range of SFH shapes, including bursty (with higher-order PCs), extended, and quiescent modes, as we have indicated by the tests in SFH space. Moreover, the PCA approach allows for efficient SSM—for instance, by combining with a Bayesian framework. Unlike the approaches mentioned above, where physical-/observational-driven priors are directly implemented in SFH models, we explore the PC coefficient distributions in Appendix C, by projecting simulated SFHs from different galaxy populations onto the PCA basis. We find that most PC coefficient distributions are Gaussian-like and exhibit significant overlap across populations. This suggests that the PCA-based models generalize well across simulations and environments, reflecting a shared underlying structure in galaxy SFHs, despite differences in physics and resolution. Such marginalized distributions can serve as a preliminary mathematical prior. Naturally, this may lead to imposing a multivariate Gaussian prior, which is often used to control the smoothness of SFH shapes. While our findings indicate broad representational capacity, the lack of explicit physical constraints (e.g., nonnegativity and smoothness) and the population-dependent shift in PC1 underscore the need for caution, particularly when extrapolating to high redshifts or observational data. Future improvements could include incorporating physically motivated priors (e.g., penalizing SFHs that may place a galaxy too far off from the main sequence of star-forming galaxies) or constraining the PCA coefficient space to ensure more physically plausible reconstructions (e.g., sampling the PC space and rejecting combinations that yield nonphysical SFHs), while preserving flexibility.

Overall, the PCA-based models demonstrate flexibility in constructing a wide range of SFH shapes. In particular, the PCA+step model also reproduces the input mock spectra with high fidelity. While its performance in spectral space is comparable to that of the stepwise model, its enhanced ability to capture the diversity of SFH shapes makes it a compelling candidate for SSM. Thus far, our tests of the PCA-based models have been primarily theoretical. In future work, we aim to apply this new framework to full SED fitting. Leveraging BIGS, we will be able to estimate the entire posterior probability distribution of SFH parameters, as well as to explore the degeneracy in the inferred SFH. In addition, allowing the dust and metallicity to vary in SSM can give us more insight into the age–dust–metallicity degeneracy with the new SFH model. Moreover, we plan to apply the new PCA-based model to high-redshift galaxies, especially with the data provided by the James Webb Space Telescope, where the limited data generally require more physically informed and robust modeling strategies. Finally, because the physically motivated eigenhistories can be mapped directly to corresponding "PC-spectra," this enables a new type of spectral template. We plan to compare these PC-based templates with those used in state-of-the-art SSM codes, such as EAZY (G. B. Brammer et al. 2008) and pPXF (M. Cappellari 2017), offering a new approach to the efficient and physically grounded modeling of galaxy spectra.

## Acknowledgments

We gratefully acknowledge Dr. Yangyao Chen for his help with data analysis and detailed explanations. The authors also want to thank Dr. Kai Wang for the insightful discussion.

*Software:* NUMPY (C. R. Harris et al. 2020), SCIPY (P. Virtanen et al. 2020), MATPLOTLIB (J. D. Hunter 2007), PANDAS (The pandas development team 2024), LMFIT (M. Newville et al. 2023), SKLEARN (F. Pedregosa et al. 2011), H5PY (A. Collette et al. 2023).

## Appendix A
## PCA-based Model Eigenhistories

In Table 2, we provide the empirical mean vector, as well as the first 10 eigenhistories of the PCA-based models. The first column is the lookback time, from the present time to $z \lesssim 6$, and the second column is the empirical mean vector from TNG100. To implement the PCA+step model, one can replace the SFR during the past 300 million yr from the PCA model with a step function.

**Table 2**
PCA-based SFH Model Base Functions

| Lookback Time (Gyr) | Mean | PC1 | PC2 | PC3 | PC4 | PC5 | PC6 | PC7 | PC8 | PC9 | PC10 |
|---|---|---|---|---|---|---|---|---|---|---|---|
| 0.000 | 2.020 | 0.112 | 0.084 | 0.090 | −0.083 | −0.085 | −0.068 | 0.070 | −0.083 | −0.079 | −0.051 |
| 0.027 | 2.032 | 0.114 | 0.087 | 0.094 | −0.085 | −0.087 | −0.069 | 0.071 | −0.083 | −0.079 | −0.048 |
| 0.055 | 2.043 | 0.115 | 0.090 | 0.098 | −0.087 | −0.090 | −0.071 | 0.073 | −0.083 | −0.079 | −0.044 |
| 0.083 | 2.056 | 0.117 | 0.094 | 0.101 | −0.089 | −0.092 | −0.073 | 0.074 | −0.083 | −0.079 | −0.041 |
| 0.112 | 2.068 | 0.118 | 0.097 | 0.105 | −0.091 | −0.095 | −0.075 | 0.076 | −0.084 | −0.079 | −0.037 |
| 0.142 | 2.079 | 0.119 | 0.100 | 0.109 | −0.093 | −0.096 | −0.075 | 0.076 | −0.082 | −0.077 | −0.033 |
| 0.173 | 2.080 | 0.120 | 0.102 | 0.110 | −0.091 | −0.094 | −0.070 | 0.070 | −0.073 | −0.070 | −0.025 |
| 0.204 | 2.081 | 0.120 | 0.104 | 0.111 | −0.090 | −0.091 | −0.064 | 0.065 | −0.063 | −0.061 | −0.017 |
| 0.237 | 2.082 | 0.121 | 0.106 | 0.113 | −0.088 | −0.088 | −0.059 | 0.059 | −0.053 | −0.053 | −0.009 |
| 0.270 | 2.084 | 0.121 | 0.109 | 0.114 | −0.086 | −0.085 | −0.053 | 0.052 | −0.043 | −0.045 | −0.000 |
| 0.304 | 2.085 | 0.122 | 0.111 | 0.115 | −0.085 | −0.083 | −0.047 | 0.046 | −0.033 | −0.036 | 0.009 |
| 0.339 | 2.086 | 0.122 | 0.113 | 0.117 | −0.083 | −0.080 | −0.041 | 0.040 | −0.022 | −0.027 | 0.018 |
| 0.376 | 2.089 | 0.123 | 0.112 | 0.113 | −0.073 | −0.067 | −0.027 | 0.022 | 0.001 | −0.004 | 0.030 |
| 0.413 | 2.092 | 0.123 | 0.112 | 0.110 | −0.064 | −0.054 | −0.012 | 0.004 | 0.024 | 0.019 | 0.042 |
| 0.451 | 2.095 | 0.123 | 0.111 | 0.106 | −0.054 | −0.040 | 0.004 | −0.014 | 0.048 | 0.044 | 0.054 |
| 0.490 | 2.098 | 0.123 | 0.109 | 0.101 | −0.043 | −0.027 | 0.018 | −0.031 | 0.068 | 0.063 | 0.064 |
| 0.530 | 2.101 | 0.123 | 0.107 | 0.094 | −0.033 | −0.014 | 0.031 | −0.042 | 0.078 | 0.072 | 0.066 |
| 0.571 | 2.104 | 0.123 | 0.104 | 0.087 | −0.022 | −0.001 | 0.044 | −0.055 | 0.089 | 0.081 | 0.069 |
| 0.613 | 2.107 | 0.123 | 0.101 | 0.080 | −0.011 | 0.013 | 0.057 | −0.067 | 0.101 | 0.090 | 0.072 |
| 0.657 | 2.110 | 0.122 | 0.098 | 0.072 | 0.001 | 0.027 | 0.071 | −0.080 | 0.112 | 0.099 | 0.075 |
| 0.701 | 2.109 | 0.122 | 0.095 | 0.064 | 0.012 | 0.039 | 0.081 | −0.088 | 0.117 | 0.101 | 0.073 |
| 0.747 | 2.105 | 0.121 | 0.091 | 0.056 | 0.022 | 0.052 | 0.088 | −0.091 | 0.115 | 0.097 | 0.067 |
| 0.794 | 2.101 | 0.120 | 0.087 | 0.048 | 0.032 | 0.064 | 0.096 | −0.095 | 0.112 | 0.093 | 0.060 |
| 0.842 | 2.105 | 0.119 | 0.083 | 0.039 | 0.043 | 0.075 | 0.102 | −0.096 | 0.106 | 0.083 | 0.051 |
| 0.892 | 2.112 | 0.119 | 0.080 | 0.030 | 0.055 | 0.085 | 0.107 | −0.096 | 0.098 | 0.071 | 0.039 |
| 0.943 | 2.120 | 0.119 | 0.076 | 0.021 | 0.068 | 0.095 | 0.113 | −0.096 | 0.090 | 0.058 | 0.027 |
| 0.995 | 2.128 | 0.119 | 0.073 | 0.012 | 0.081 | 0.106 | 0.119 | −0.096 | 0.081 | 0.045 | 0.015 |
| 1.049 | 2.130 | 0.119 | 0.068 | 0.003 | 0.090 | 0.113 | 0.119 | −0.088 | 0.067 | 0.027 | 0.002 |
| 1.104 | 2.131 | 0.118 | 0.063 | −0.007 | 0.099 | 0.118 | 0.117 | −0.077 | 0.050 | 0.007 | −0.013 |
| 1.161 | 2.136 | 0.117 | 0.058 | −0.017 | 0.109 | 0.122 | 0.113 | −0.064 | 0.031 | −0.014 | −0.026 |
| 1.219 | 2.149 | 0.117 | 0.055 | −0.028 | 0.119 | 0.125 | 0.104 | −0.047 | 0.007 | −0.038 | −0.038 |
| 1.279 | 2.163 | 0.117 | 0.051 | −0.039 | 0.129 | 0.127 | 0.095 | −0.030 | −0.018 | −0.063 | −0.050 |
| 1.340 | 2.177 | 0.118 | 0.047 | −0.050 | 0.140 | 0.130 | 0.085 | −0.012 | −0.044 | −0.088 | −0.062 |
| 1.403 | 2.190 | 0.117 | 0.042 | −0.060 | 0.145 | 0.125 | 0.069 | 0.012 | −0.064 | −0.106 | −0.064 |
| 1.468 | 2.204 | 0.117 | 0.037 | −0.071 | 0.149 | 0.120 | 0.054 | 0.035 | −0.085 | −0.124 | −0.066 |
| 1.534 | 2.204 | 0.116 | 0.030 | −0.078 | 0.149 | 0.109 | 0.033 | 0.054 | −0.097 | −0.126 | −0.057 |
| 1.603 | 2.203 | 0.115 | 0.024 | −0.087 | 0.148 | 0.099 | 0.012 | 0.074 | −0.109 | −0.128 | −0.047 |
| 1.673 | 2.204 | 0.114 | 0.017 | −0.094 | 0.147 | 0.087 | −0.009 | 0.092 | −0.118 | −0.126 | −0.034 |
| 1.745 | 2.213 | 0.112 | 0.011 | −0.100 | 0.139 | 0.068 | −0.031 | 0.103 | −0.112 | −0.103 | −0.012 |
| 1.819 | 2.222 | 0.111 | 0.004 | −0.106 | 0.132 | 0.048 | −0.053 | 0.114 | −0.106 | −0.078 | 0.012 |
| 1.894 | 2.227 | 0.109 | −0.002 | −0.112 | 0.125 | 0.029 | −0.075 | 0.122 | −0.095 | −0.052 | 0.033 |
| 1.972 | 2.229 | 0.107 | −0.009 | −0.118 | 0.120 | 0.010 | −0.096 | 0.127 | −0.079 | −0.023 | 0.052 |
| 2.052 | 2.250 | 0.107 | −0.014 | −0.124 | 0.114 | −0.009 | −0.113 | 0.128 | −0.057 | 0.008 | 0.077 |
| 2.135 | 2.274 | 0.108 | −0.019 | −0.130 | 0.108 | −0.029 | −0.131 | 0.128 | −0.033 | 0.041 | 0.104 |
| 2.219 | 2.291 | 0.107 | −0.025 | −0.135 | 0.100 | −0.051 | −0.144 | 0.118 | −0.005 | 0.072 | 0.121 |
| 2.306 | 2.303 | 0.106 | −0.030 | −0.137 | 0.090 | −0.075 | −0.152 | 0.101 | 0.027 | 0.098 | 0.127 |
| 2.395 | 2.311 | 0.104 | −0.036 | −0.137 | 0.076 | −0.092 | −0.147 | 0.075 | 0.057 | 0.107 | 0.112 |
| 2.486 | 2.319 | 0.103 | −0.041 | −0.137 | 0.060 | −0.109 | −0.142 | 0.049 | 0.087 | 0.116 | 0.095 |
| 2.580 | 2.338 | 0.102 | −0.046 | −0.136 | 0.046 | −0.125 | −0.129 | 0.018 | 0.109 | 0.114 | 0.076 |
| 2.676 | 2.355 | 0.100 | −0.051 | −0.135 | 0.031 | −0.140 | −0.114 | −0.015 | 0.130 | 0.111 | 0.055 |
| 2.775 | 2.341 | 0.097 | −0.059 | −0.134 | 0.017 | −0.145 | −0.094 | −0.047 | 0.143 | 0.096 | 0.026 |
| 2.877 | 2.361 | 0.096 | −0.066 | −0.135 | 0.001 | −0.155 | −0.067 | −0.076 | 0.145 | 0.070 | −0.009 |
| 2.981 | 2.386 | 0.096 | −0.074 | −0.136 | −0.015 | −0.165 | −0.040 | −0.104 | 0.144 | 0.040 | −0.044 |
| 3.089 | 2.390 | 0.093 | −0.081 | −0.131 | −0.031 | −0.167 | −0.015 | −0.122 | 0.133 | −0.004 | −0.072 |
| 3.199 | 2.410 | 0.092 | −0.088 | −0.126 | −0.048 | −0.166 | 0.015 | −0.137 | 0.115 | −0.049 | −0.091 |
| 3.312 | 2.437 | 0.091 | −0.093 | −0.120 | −0.065 | −0.156 | 0.048 | −0.142 | 0.084 | −0.087 | −0.092 |
| 3.428 | 2.437 | 0.088 | −0.097 | −0.110 | −0.077 | −0.139 | 0.081 | −0.138 | 0.040 | −0.116 | −0.081 |
| 3.547 | 2.457 | 0.086 | −0.102 | −0.102 | −0.091 | −0.121 | 0.116 | −0.133 | −0.004 | −0.143 | −0.068 |
| 3.669 | 2.500 | 0.084 | −0.109 | −0.096 | −0.104 | −0.104 | 0.146 | −0.123 | −0.045 | −0.152 | −0.047 |
| 3.795 | 2.516 | 0.080 | −0.115 | −0.086 | −0.115 | −0.086 | 0.168 | −0.107 | −0.081 | −0.143 | −0.017 |
| 3.924 | 2.533 | 0.079 | −0.117 | −0.074 | −0.124 | −0.068 | 0.183 | −0.080 | −0.115 | −0.125 | 0.031 |
| 4.057 | 2.545 | 0.075 | −0.118 | −0.063 | −0.129 | −0.046 | 0.188 | −0.047 | −0.135 | −0.093 | 0.079 |
| 4.193 | 2.541 | 0.072 | −0.121 | −0.051 | −0.133 | −0.020 | 0.186 | −0.002 | −0.146 | −0.043 | 0.123 |

**Table 2**
(Continued)

| Lookback Time (Gyr) | Mean | PC1 | PC2 | PC3 | PC4 | PC5 | PC6 | PC7 | PC8 | PC9 | PC10 |
|---|---|---|---|---|---|---|---|---|---|---|---|
| 4.333 | 2.565 | 0.069 | −0.125 | −0.039 | −0.136 | 0.006 | 0.182 | 0.045 | −0.148 | 0.014 | 0.157 |
| 4.476 | 2.597 | 0.067 | −0.129 | −0.027 | −0.136 | 0.032 | 0.173 | 0.091 | −0.136 | 0.070 | 0.170 |
| 4.624 | 2.627 | 0.064 | −0.132 | −0.013 | −0.134 | 0.058 | 0.153 | 0.132 | −0.105 | 0.116 | 0.159 |
| 4.775 | 2.678 | 0.061 | −0.136 | 0.001 | −0.134 | 0.082 | 0.121 | 0.166 | −0.059 | 0.152 | 0.126 |
| 4.931 | 2.661 | 0.056 | −0.134 | 0.016 | −0.124 | 0.105 | 0.091 | 0.188 | −0.005 | 0.164 | 0.072 |
| 5.091 | 2.743 | 0.055 | −0.139 | 0.029 | −0.127 | 0.129 | 0.063 | 0.216 | 0.052 | 0.169 | 0.010 |
| 5.255 | 2.766 | 0.050 | −0.143 | 0.041 | −0.125 | 0.143 | 0.030 | 0.223 | 0.107 | 0.140 | −0.052 |
| 5.423 | 2.796 | 0.047 | −0.144 | 0.054 | −0.110 | 0.148 | −0.012 | 0.189 | 0.156 | 0.072 | −0.102 |
| 5.596 | 2.802 | 0.042 | −0.141 | 0.066 | −0.090 | 0.143 | −0.050 | 0.142 | 0.183 | −0.010 | −0.124 |
| 5.773 | 2.835 | 0.039 | −0.139 | 0.079 | −0.072 | 0.144 | −0.083 | 0.092 | 0.196 | −0.086 | −0.117 |
| 5.956 | 2.898 | 0.034 | −0.138 | 0.095 | −0.052 | 0.141 | −0.115 | 0.036 | 0.202 | −0.161 | −0.091 |
| 6.143 | 2.958 | 0.028 | −0.140 | 0.105 | −0.027 | 0.136 | −0.141 | −0.021 | 0.196 | −0.224 | −0.037 |
| 6.335 | 3.023 | 0.024 | −0.138 | 0.110 | −0.008 | 0.121 | −0.148 | −0.068 | 0.144 | −0.228 | 0.039 |
| 6.533 | 3.036 | 0.016 | −0.137 | 0.115 | 0.006 | 0.097 | −0.141 | −0.106 | 0.081 | −0.188 | 0.115 |
| 6.736 | 3.082 | 0.009 | −0.135 | 0.122 | 0.022 | 0.084 | −0.138 | −0.134 | 0.024 | −0.152 | 0.182 |
| 6.944 | 3.143 | −0.000 | −0.132 | 0.136 | 0.047 | 0.064 | −0.134 | −0.161 | −0.038 | −0.113 | 0.241 |
| 7.158 | 3.170 | −0.006 | −0.121 | 0.137 | 0.067 | 0.037 | −0.116 | −0.163 | −0.093 | −0.046 | 0.242 |
| 7.378 | 3.204 | −0.015 | −0.115 | 0.142 | 0.092 | 0.008 | −0.098 | −0.167 | −0.149 | 0.031 | 0.228 |
| 7.603 | 3.216 | −0.022 | −0.104 | 0.138 | 0.108 | −0.026 | −0.071 | −0.143 | −0.162 | 0.090 | 0.142 |
| 7.835 | 3.225 | −0.033 | −0.090 | 0.140 | 0.127 | −0.059 | −0.036 | −0.115 | −0.165 | 0.144 | 0.058 |
| 8.073 | 3.237 | −0.040 | −0.076 | 0.140 | 0.146 | −0.091 | −0.002 | −0.083 | −0.154 | 0.176 | −0.040 |
| 8.317 | 3.242 | −0.045 | −0.067 | 0.130 | 0.145 | −0.091 | 0.023 | −0.045 | −0.103 | 0.149 | −0.100 |
| 8.568 | 3.252 | −0.056 | −0.054 | 0.120 | 0.143 | −0.105 | 0.057 | 0.000 | −0.056 | 0.113 | −0.138 |
| 8.826 | 3.300 | −0.072 | −0.036 | 0.116 | 0.155 | −0.135 | 0.100 | 0.051 | −0.011 | 0.074 | −0.181 |
| 9.090 | 3.293 | −0.083 | −0.017 | 0.101 | 0.149 | −0.144 | 0.118 | 0.095 | 0.045 | 0.017 | −0.141 |
| 9.362 | 3.272 | −0.095 | 0.001 | 0.085 | 0.142 | −0.150 | 0.138 | 0.130 | 0.094 | −0.045 | −0.109 |
| 9.641 | 3.259 | −0.104 | 0.022 | 0.065 | 0.132 | −0.150 | 0.146 | 0.140 | 0.111 | −0.077 | −0.007 |
| 9.927 | 3.227 | −0.110 | 0.041 | 0.045 | 0.114 | −0.139 | 0.146 | 0.139 | 0.124 | −0.098 | 0.093 |
| 10.222 | 3.176 | −0.123 | 0.074 | 0.014 | 0.085 | −0.125 | 0.143 | 0.141 | 0.141 | −0.149 | 0.222 |
| 10.524 | 3.097 | −0.130 | 0.106 | −0.020 | 0.040 | −0.085 | 0.116 | 0.129 | 0.143 | −0.172 | 0.265 |
| 10.834 | 3.007 | −0.141 | 0.130 | −0.054 | 0.004 | −0.051 | 0.086 | 0.101 | 0.121 | −0.145 | 0.254 |
| 11.153 | 2.894 | −0.152 | 0.174 | −0.110 | −0.054 | 0.004 | 0.038 | 0.050 | 0.097 | −0.122 | 0.228 |
| 11.480 | 2.716 | −0.157 | 0.221 | −0.173 | −0.136 | 0.079 | −0.035 | −0.028 | 0.023 | −0.042 | 0.130 |
| 11.816 | 2.398 | −0.141 | 0.223 | −0.193 | −0.176 | 0.130 | −0.091 | −0.091 | −0.048 | 0.040 | 0.004 |
| 12.161 | 1.935 | −0.116 | 0.198 | −0.184 | −0.188 | 0.158 | −0.129 | −0.135 | −0.103 | 0.108 | −0.103 |
| 12.515 | 1.255 | −0.069 | 0.123 | −0.120 | −0.129 | 0.113 | −0.095 | −0.099 | −0.084 | 0.087 | −0.104 |
| 12.879 | 0.502 | −0.028 | 0.048 | −0.046 | −0.050 | 0.044 | −0.039 | −0.041 | −0.037 | 0.038 | −0.048 |

**Note.** The lookback time, shown in the first column, is sampled in a logarithmic scale (base 10). The following columns are the empirical mean vector, as well as the first 10 eigenhistories, both in linear scale. The PCA+step model uses the same empirical mean vector and eigenhistories here.

## Appendix B
## Test Results for Using Different Numbers of Eigenhistories

In Section 3, we present the PVE test of the PCA-based models. There, we find that the cumulative data variance explained by the first few eigenhistories increases the fastest and that the first 5 eigenhistories can capture ∼65% of the data variances. Here, we study the effects of using a different number of eigenhistories quantitatively, in both the SFH and spectral spaces. The results obtained from the entire sample of 9168 TNG100 central galaxies are shown in Figure 11. In both panels, we show the median $\Delta$ values using solid lines and the 25%–75% percentile values using dashed lines.

In SFH space (left panel), it is seen that as the number of eigenhistories used in each model increases, the median $\Delta_{\rm SFH}$ value gradually decreases. This is expected, since when more higher-order eigenhistories are used, more small-scale details can be recovered, thus making the best-fit SFHs closer to the input SFHs. Similar to the PVE test, within the first few eigenhistories, the median $\Delta_{\rm SFH}$ curves are steeper. The similar performances of the two PCA-based models are also expected.

On the contrary, the statistical performance becomes more diverse in spectral space, as shown in the right panel of Figure 11. Again, as the number of eigenhistories increases, the spectra generated by the best-fit SFHs better match the input mock spectra generated by the input SFHs. However, when comparing the models with the same number of eigenhistories, the median $\Delta_{\rm spec}$ values of the PCA+step model are roughly an order of magnitude smaller. Clearly, the effects on the spectral shape come from the most recent star formation (within the past ∼0.3 Gyr). Furthermore, the decreasing trends become much shallower after the first five eigenhistories. This is because the higher-order components govern the high-frequency details of the SFHs, which are not well reflected in the spectra. In particular, from the PCA+step model, the median $\Delta_{\rm spec}$ becomes quite flat, starting from the fifth eigenhistory. Thus, we argue that using the first five eigenhistories in our PCA-based models is sufficient to model the overall properties of galaxy spectra.

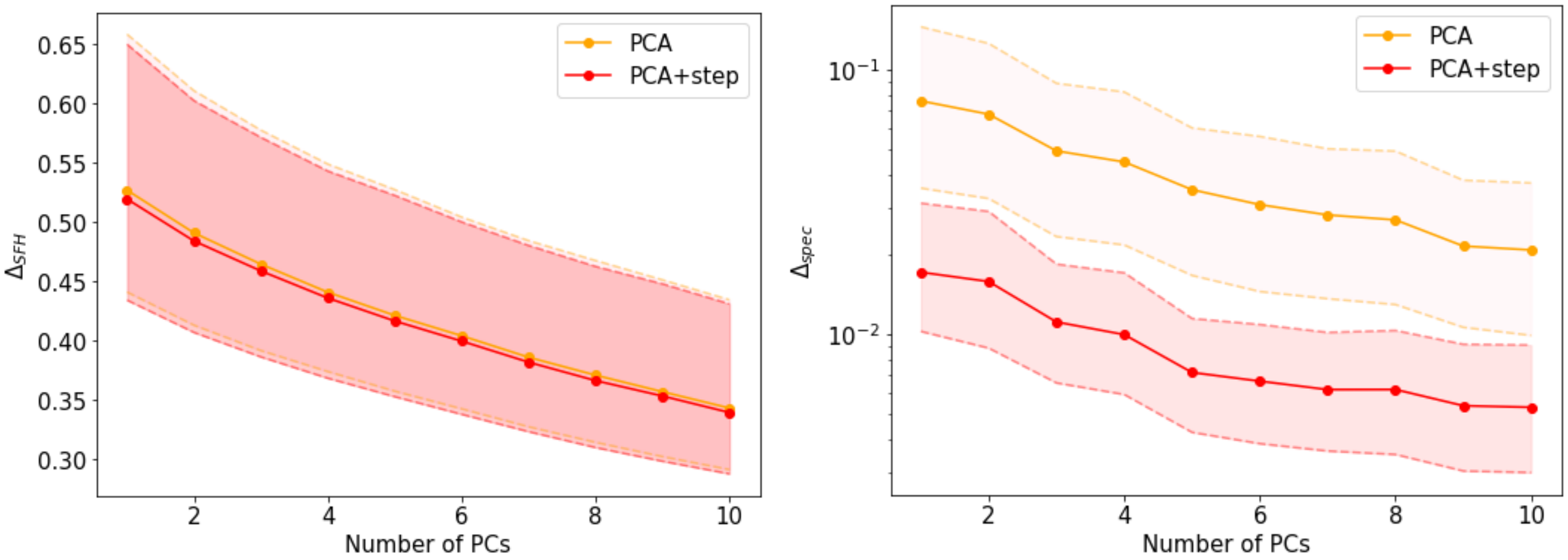

**Figure 11.** The $\Delta$ values of the PCA model (orange) and the PCA+step model (red) obtained by using the first one to 10 eigenhistories. The solid curves mark the median $\Delta$ values, while the dashed curves mark the 25% and 75% percentile values. The left and right panels show $\Delta_{\rm SFH}$ and $\Delta_{\rm spec}$, respectively.

# Appendix C PC Space Coefficient Distributions

To assess the generality and representational capacity of the PCA-based SFH models, we examine the distributions of PC coefficients obtained by directly projecting SFHs from different galaxy populations onto the PCA basis (i.e., the eigenhistories). In Figure 12, we show a 2D corner plot for the first 10 coefficient distributions. The contours represent 0.393, 0.675, and 0.864 confidence levels, corresponding to $1\sigma$, $1.5\sigma$, and $2\sigma$ separations from the center of a 2D Gaussian distribution. These coefficient distributions can be interpreted as mathematical priors, reflecting the intrinsic diversity of the SFHs encoded in the simulations before any spectral fitting is performed. We find that for most PCs, the projected coefficients follow approximately Gaussian distributions across all populations considered, i.e., TNG centrals, TNG satellites, and EAGLE centrals. This Gaussian-like behavior indicates that the SFHs from these simulations, despite their different formation environments and physical prescriptions, occupy a relatively well-defined and smooth region of PCA space.

A particularly notable result is that the coefficient distributions from different populations largely overlap in most PCs. This overlap suggests that the PCA basis derived from the TNG central galaxies is general enough to represent the SFHs of a broader range of systems, including satellite galaxies and galaxies from an entirely different simulation, EAGLE. Such overlap supports the idea that while the PCA approach is simulation-specific, the dominant modes of SFH variability captured by the PCA-based models are not overly tailored to a single set of conditions and may reflect more universal trends in galaxy evolution.

Meanwhile, we observe differences in the spreads of the distributions. The TNG satellite galaxies show broader coefficient distributions than either the TNG or EAGLE centrals. This increased spread is likely due to the greater variety of evolutionary paths experienced by the satellites, including environmental effects like quenching, stripping, and interactions, which introduce more diversity into their SFHs. In contrast, the EAGLE central galaxies exhibit narrower distributions than TNG centrals, despite the PCA-based models being trained on the latter sample and the differences in the subgrid physics implemented in the two simulations. This narrower spread may indicate that the EAGLE centrals follow more homogeneous SFH trajectories, or that the PCA basis, though derived from TNG centrals, still provides an efficient representation of the EAGLE SFHs within a compact coefficient range.

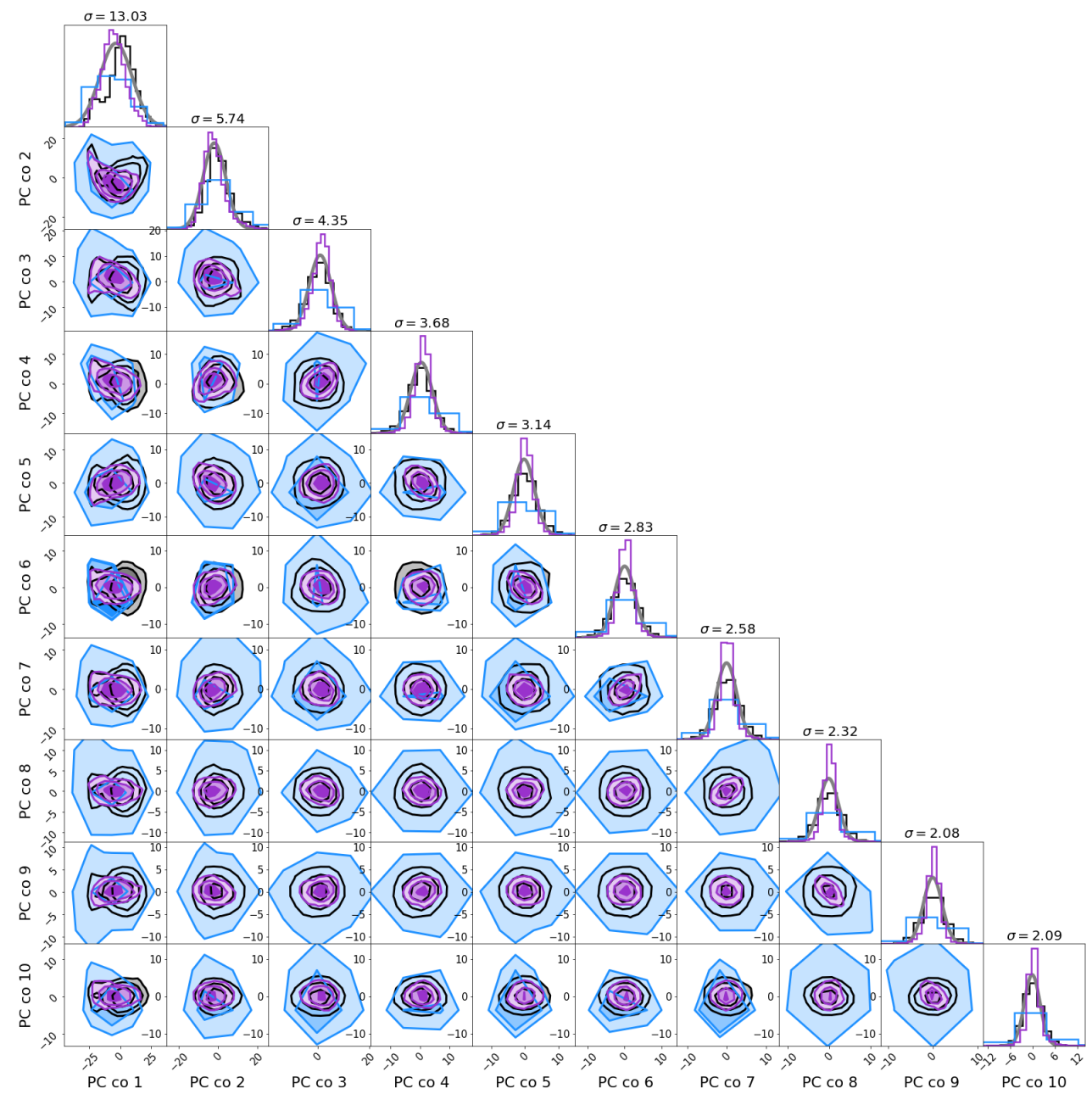

**Figure 12.** 2D coefficient distribution of the three galaxy populations (TNG central: black; TNG satellite: blue; and EAGLE central: purple) in PC space for the first 10 eigenhistories. The contours represent $1\sigma$, $1.5\sigma$, and $2\sigma$ separations from the center of the distribution. Due to the loose distribution of the TNG satellite population, the inner contours are often not plotted in the figures. The gray curves in the 1D histogram plots are the best-fit Gaussian curves for the three populations combined, and the dispersion of each curve is listed in the title.

The only major divergence occurs in the distribution of PC1, which represents the dominant mode of SFH variation from the mean vector, often associated with early versus late star formation activities or quenching timescales. Here, we see a clear offset between the populations. This offset is not unexpected, since PC1 captures the largest structural variance in SFH shapes from the mean vector obtained from the training data, and it is sensitive to intrinsic variations between simulations. Such differences may reflect genuine physical distinctions, such as earlier quenching in TNG simulations or more extended star formation in EAGLE, rather than a failure of the PCA representation itself.

We combine the projected coefficients from the three populations and fit them with a Gaussian. The best-fit Gaussian distribution curves are plotted in gray at each 1D histogram plot, with the dispersion listed on top. These Gaussian curves center at zero, and as we go to higher-order PCs, the dispersion gets smaller. Such behaviors are highly expected, since lower-order PCs contain the most significant variance from the training data. Furthermore, the near-perfect Gaussian fit of the coefficient distributions indicates uncorrelated parameters, which is one of the significances of the PCA approach, i.e., the construction of the orthogonal basis.

## ORCID iDs

Yanzhe Zhang https://orcid.org/0000-0002-5564-0254
H. J. Mo https://orcid.org/0000-0001-5356-2419
Katherine E. Whitaker https://orcid.org/0000-0001-7160-3632
Shuang Zhou https://orcid.org/0000-0002-8999-6814